\documentclass[submission,Phys]{SciPost}
\usepackage{color}
\usepackage{amsfonts}
\usepackage{amsmath}
\usepackage{subfig}
\usepackage{amssymb}
\usepackage{amsthm}
\usepackage{amscd}
\usepackage{eucal}
\usepackage{bm}
\usepackage{booktabs}
\usepackage{graphicx}
\usepackage{float}
\input{macro.lib}
\definecolor{dgreen}{rgb}{0.2 ,0.54, 0.2}

\begin{document}
\begin{center}{\Large \textbf{
Soliton confinement in  the double sine-Gordon model
}}\end{center}

\begin{center}
S.~B.~Rutkevich
\end{center}
\begin{center}
Fakult\"at f\"ur Mathematik und Naturwissenschaften, Bergische Universit\"at Wuppertal, 42097 Wuppertal, Germany
\end{center}

\begin{center}
\date{December 23, 2023}
\end{center}
\section*{Abstract}
The double sine-Gordon field theory in the weak confinement regime is studied. It represents the small 
non-integrable deformation of the standard sine-Gordon model  caused by the cosine perturbation with
the frequency reduced by the factor of 2. 
This perturbation leads to the 
confinement of the sine-Gordon solitons, which become coupled into the 'meson' bound states.  
We classify the meson states  in the weak confinement regime, and obtain three asymptotic expansions for their masses, which can  be used in  different regions of the model parameters. It is shown, that the sine-Gordon breathers, 
slightly deformed by the perturbation term, transform into the mesons upon increase of the sine-Gordon coupling constant.

\vspace{10pt}
\noindent\rule{\textwidth}{1pt}
\tableofcontents\thispagestyle{fancy}
\noindent\rule{\textwidth}{1pt}
\vspace{10pt}
\section{Introduction}
Integrable models of statistical mechanics and quantum field theory \cite{Bax,Mussardo10} play an important role 
in the present understanding of the  properties of the condense matter systems
in the critical region close to the point of the continuous phase transition. Due to the universality 
of critical fluctuations, the exact solution of such a model provides  a mathematically reliable information 
not only about the  integrable model itself, but also about the whole universality class it represents.

It turns out, however, that a number of interesting physical phenomena, such as  particle decay, 
nucleation of  domains of the equilibrium phase in the false vacuum surrounding, confinement of topological excitations, etc., 
require non-integrable  models for their description.  If the proper  non-integrable model can be viewed as a 
small deformation of some integrable one,  perturbative expansions around the integrable point could provide 
a useful insight into the  realm of physical phenomena  described by such a non-integrable model \cite{Mussardo10,Mussardo_2011}.  This idea was first realized  in the form factor perturbation theory (FFPT), which was introduced by Delfino, Mussardo, and Simonetti \cite{Del96}.

In this paper we study the non-integrable deformation of the sine-Gordon field theory, which is known in the literature 
as the two-frequency or double sine-Gordon (dsG) model  \cite{Cam86,Del98,Baj2001,Mus_2004,TAKACS2006353}. We address the particular case of this model, which is specified by the 
Euclidean action:
\begin{equation}\label{psG0}
\mathcal{A}_{dsG}=\int d^2 x  \left[
\frac{1}{16 \pi} (\partial_\mathfrak{a} \varphi)^2+V(\varphi)
\right],
\end{equation}
where $\mathfrak{a} =0,1$, and 
\begin{equation}\label{V0}
V(\varphi)=-2 \mu \cos (\beta \varphi)- 2\Lambda \cos(\beta \varphi/2).
\end{equation}
The parameters $\mu, \beta$, and $\Lambda$ are positive, and $0<\beta\le 1$. Along with the sine-Gordon coupling constant 
$\beta$, the
so-called renormalized coupling constant $\xi\in(0,+\infty)$, which is related with $\beta$ as
  \begin{equation} \label{xi}
   \xi=\frac{\beta^2}{1-\beta^2},
  \end{equation}
is also widely used.

At $\Lambda=0$,  action \eqref{psG0} describes the standard  integrable sine-Gordon  (sG) model. The description
of its various properties can be found, {\it{e.g.}} in the monograph \cite{Mussardo10} and references therein.
At $\Lambda>0$, the second 
term on the right-hand side of \eqref{V0} breaks  integrability of the model.  It was shown by Delfino and Mussardo  \cite{Del98}, 
that this perturbation term induces
a linear attractive potential between the solitons of the sG model leading to their confinement:
isolated solitons do not exist anymore in the system, and two  solitons bind into compound particles - the `mesons'.
Recently, the $\Lambda$-dependencies  of the mass of the lightest meson
at several fixed values of the parameter $\beta$ in model \eqref{psG0}, \eqref{V0} were studied numerically by Roy and Lukyanov \cite{Roy2023}.

The aim of the present work is to  classify the meson states in the  dsG model \eqref{psG0}  in the weak confinement regime  at 
small $\Lambda\to+0$, and to calculate the meson mass spectra perturbatively in this small parameter. We obtain several initial terms of three asymptotic expansions, which describe the meson masses 
for   $\xi\in (0,+\infty)$. It is shown, in particular, that at any  small $\Lambda>0$, 
there is no qualitative difference between the  sG breathers, which survive at $\Lambda=0$ for 
$\xi<1$, and the newly formed mesons, which are present only  at $\Lambda>0$. At a fixed small $\Lambda>0$, the  breather,
slightly deformed by the  perturbation $\sim\Lambda$,
smoothly transforms upon  increase of the parameter $\xi$ into the newly formed meson state.

The rest of the paper is organised as follows. In Section \ref{Sec2} we recall some well-known properties of the sine-Gordon
field theory and its  two-frequency deformation defined by \eqref{psG0}, \eqref{V0}. Section \ref{Sec:WCE} contains a  brief 
review of two  perturbative  techniques, which were recently developed for the calculation of the meson mass spectra in  several  QFT and spin-chain models in the weak confinement regime. Applying one of these techniques to the dsG model 
\eqref{psG0} at a small $\Lambda>0$, we obtain two asymptotic expansions for the meson masses, which hold
in different regions of the meson parameters. In Section \ref{Sec_LEE} we derive the third asymptotic expansion, 
which determines  the masses of light mesons  for $\xi$ in the 
crossover intervals close to the inverse natural numbers $\xi_n=1/n$. Together, three asymptotic formulas
obtained in  Sections \ref{Sec:WCE} and  \ref{Sec_LEE}    determine  
the evolution of  masses of all meson modes upon tuning the parameter $\xi$ in the whole range of its variation
$0<\xi<\infty$. This evolution is clarified and illustrated  in Section \ref{Sec:Ev}. Concluding remarks are presented in
Section \ref{Sec:Conc}. Finally, Appendix~\ref{LibLin} contains the calculation of the meson energy spectra in the deformed
Lieb-Liniger model. The results of this appendix are used in Section~\ref{Sec_LEE}.
\section{Sine-Gordon model and its deformation \label{Sec2}}
In this section we  remind the reader some well-known facts about  the pure SG model and its deformation \eqref{psG0}, \eqref{V0}.

\begin{figure}[htb]
\includegraphics[width=\linewidth, angle=00]{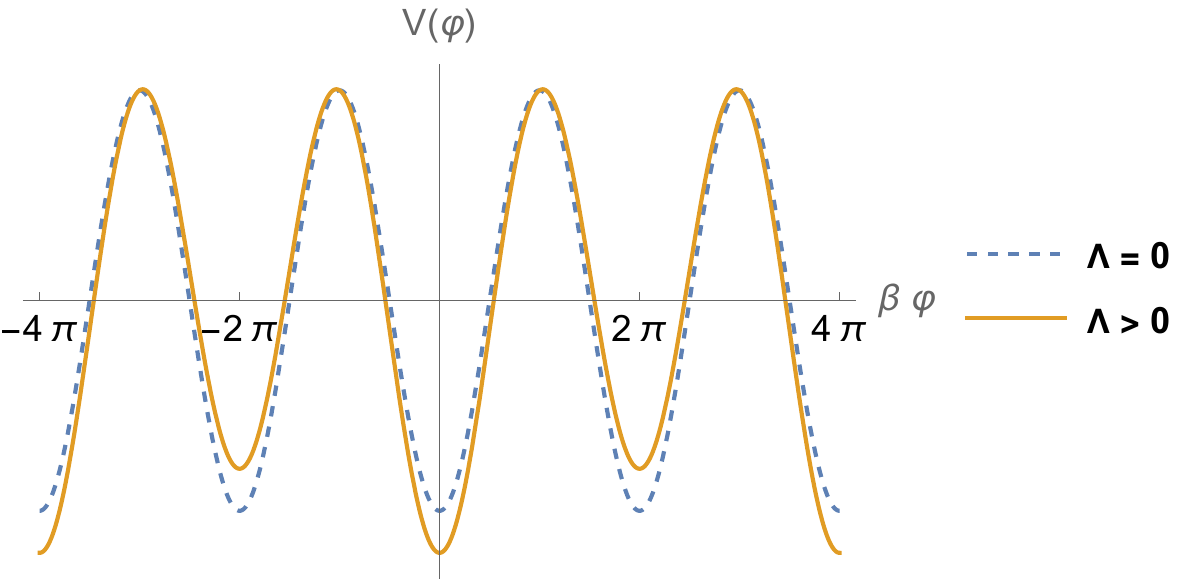}
\caption{The potential \eqref{V0} at $\Lambda=0$ (dashed curve), and at a small $\Lambda>0$ (solid curve).
\label{fig:Vph} } 
\end{figure}
  
 The profiles  of the potential $V(\varphi)$ at $\Lambda=0$, and at a small $\Lambda>0$ are shown in Figure \ref{fig:Vph} 
 by the dashed, and solid curves, respectively. 
At $\Lambda=0$, the potential \eqref{V0} is the $2\pi/\beta$-periodical function of $\varphi$, which takes its minimum
value 
\begin{equation}\label{Vm}
V_{min}(\mu,\Lambda=0)=-2 \mu,  
\end{equation}
at $\varphi=2 \pi l/\beta$, with $l\in \mathbb{Z}$. At $\Lambda\ne 0$, the period of the potential
$V(\varphi)$ increases by the factor of 2.  At a small enough  $\Lambda\in (0,1/4)$, the locations of minima remain the same, $\varphi=2 \pi l/\beta$,
while the minimum values of the potential $V(\varphi)$ corresponding to even and odd $l$ become shifted downwards, and upwards, respectively: 
\begin{equation}
V_{min}(\mu,\Lambda,l)=\begin{cases}
-2 \mu-2\Lambda,&\text{for  even } l,\\
-2 \mu+2\Lambda, &\text{for  odd }  l.
\end{cases}
\end{equation}

Due to the $4\pi/\beta$-periodicity of the potential \eqref{V0}, it is natural to compactify the scalar field variable $\varphi$ 
by putting  it on the 
circle with the circumference $4\pi/\beta$:
\[
\varphi\sim \varphi+ \frac{4\pi l}{\beta}, \text {with } l\in \mathbb{Z}.
\]
Upon this identification, the model \eqref{psG0} reduces at $\Lambda=0$ to the two-folded sine-Gordon model $\mathrm{sG}(\beta,2)$ \cite{Baj2000}.
In the infinite volume, this model has two energetically degenerate vacuums $|vac\rangle^{(a)}$, $a=0,1$, with the 
energy density $\mathcal{E}_{sG}$. In the {\it repulsive regime} at $\xi>1$, the particle sector is represented solely by the kinks (solitons and antisolitons)
$|K_{ab}(\alpha)\rangle_s$ interpolating between the vacua  $a$ and $b$. In the {\it attractive regime} at $\xi\in (0,1)$, the two-kink bound states
(the breathers) are also allowed at $\Lambda=0$ \cite{Zam77}. The kink state $|K_{ab}(\alpha)\rangle_s$
is characterized by the rapidity $\alpha\in \mathbb{R}$, 
and by the isospin\begin{footnote}
{The origin of the $O(2)$ isotopic symmetry of the sG model was discussed in \cite{ZZ79} in page 268.}
\end{footnote} (topological charge) $s$, which takes the value $+1/2$ for solitons, and $-1/2$ for antisolitons. The energy $\omega$ and the momentum $p$ of a kink can be  parametrized by its rapidity $\alpha$ 
in the usual way:
\begin{equation}\label{DL}
\omega(\alpha)= m \cosh \alpha, \quad p(\alpha)= m \sinh \alpha,
\end{equation}
where $m$ is the kink mass.
 The dimensional coupling constant $\mu$ in the sG potential \eqref{V0} is related to the kink mass $m$ as 
 follows:
 \begin{equation}\label{mum}
 \mu =\kappa(\xi) \,m^{2/(\xi+1)}, 
 \end{equation}
    where $\xi$ is the renormalized coupling constant \eqref{xi},
 and the constant $\kappa(\xi)$ is known due to  Al.~B.~Zamolodchikov \cite{Al_Z95}:
\begin{equation}
  \kappa(\xi)=\frac{1}{\pi}\frac{\Gamma\left(\frac{\xi}{\xi+1}\right)}{\Gamma\left(\frac{1}{\xi+1}\right)}
  \left[
  \frac{\sqrt{\pi}\,\Gamma\left(\frac{\xi+1}{2}\right)}{2\,\Gamma\left(\frac{\xi}{2}\right)}
  \right]^{2/(\xi+1)}.
\end{equation}

\begin{figure}[htb]
\includegraphics[width=\linewidth, angle=00]{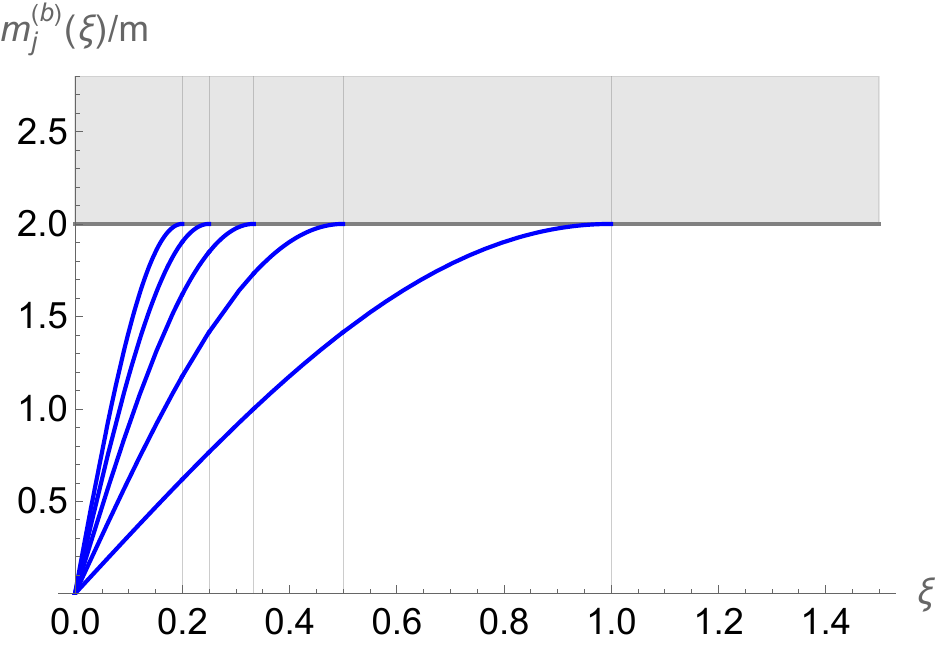}

\caption{The $\xi$-dependencies due to  \eqref{mbr} of the masses of five lightest breathers.
\label{fig:Br} } 
\end{figure}

In the attractive regime at $\xi\in \left(\frac{1}{j+1},\frac{1}{j}\right)$, there are $j$ breather states with masses \cite{KorFad75,Zam77}: 
\begin{equation}\label{mbr}
m_{j}^{(b)}(\xi)=2 m \sin\left(\frac{\pi j\xi}{2}\right). 
\end{equation}
We will distinguish the breathers with odd and even $j$ by the breather parity:
\begin{equation}\label{brs}
\iota(j)=(-1)^j.
\end{equation}
The variation of the breather masses with $\xi$ for five lightest breather is shown in Figure \ref{fig:Br}.
The continuous spectrum in the  two-soliton sector, which is shown in grey in this  figure, fills the energies $E>2m$.
   
 The two-kink excitations in the two-folded sine-Gordon model $\mathrm{sG}(\beta,2)$  are represented by the states 
 $|K_{ab}(\alpha_1)K_{ba}(\alpha_2)\rangle_{s_1 s_2}$. 
 We shall use also the two-kink states $ |K_{ab}(\alpha_1)K_{ba}(\alpha_2)\rangle_{\pm}$ defined by the relations:
 \begin{align}
 |K_{ab}(\alpha_1)K_{ba}(\alpha_2)\rangle_{\pm}=
 \frac{1}{\sqrt{2}}[ |K_{ab}(\alpha_1)K_{ba}(\alpha_2)\rangle_{1/2,-1/2}\\\nonumber
 \pm  |K_{ab}(\alpha_1)K_{ba}(\alpha_2)\rangle_{-1/2,1/2}].
 \end{align}
 
The two-kink scattering matrix in the sG model was found by  A.~B.~Zamolodchikov \cite{Zam77}. Its 
representation in the $\mathrm{sG}(\beta,2)$  model can be described by the commutation relations:
\begin{subequations}\label{FZC}
\begin{align}
&|K_{ab}(\alpha_1)K_{ba}(\alpha_2)\rangle_{1/2,1/2}=S_0(\alpha_1-\alpha_2)
|K_{ab}(\alpha_2)K_{ba}(\alpha_1)\rangle_{1/2,1/2}, \\
&|K_{ab}(\alpha_1)K_{ba}(\alpha_2)\rangle_{-1/2,-1/2}=S_0(\alpha_1-\alpha_2)
|K_{ab}(\alpha_2)K_{ba}(\alpha_1)\rangle_{-1/2,-1/2}, \\
&|K_{ab}(\alpha_1)K_{ba}(\alpha_2)\rangle_{\pm}=S_\pm(\alpha_1-\alpha_2)
|K_{ab}(\alpha_2)K_{ba}(\alpha_1)\rangle_{\pm},
\end{align}
\end{subequations}
where 
\begin{equation}\label{Siota}
S_\iota(\alpha,\xi)=-e^{i \theta_\iota(\alpha,\xi)}, 
\end{equation}
with $\iota=0,\pm$. 
The scattering phase $\theta_0(\alpha)$ is given by the integral formula:
\begin{equation}\label{th0}
\theta_0(\alpha,\xi)=-\int_0^\infty \frac{dx}{ x} \, \frac{\sinh\left(\frac{x}{2\xi}-\frac{x}{2}\right)  \sin\left( \frac{\alpha x}{\pi \xi}\right)}
{\sinh(x/2)\cosh\left(\frac{x}{2\xi}\right) }.
\end{equation}
The scattering phases $\theta_\pm(\alpha,\xi)$ are determined by formulas:
\begin{subequations}\label{thet}
\begin{align}\label{Sp}
&\theta_\pm(\alpha,\xi)=\chi_\pm(\alpha)+\theta_0(\alpha,\xi),\\\label{Spp}
& \chi_+(\alpha,\xi)=\arg\left[\frac{\sinh[(i \pi +\alpha)/(2 \xi)]}{\sinh[(i \pi -\alpha)/(2 \xi)]}\right],\\
&\chi_-(\alpha,\xi)=\arg\left[\frac{\cosh[(i \pi +\alpha)/(2 \xi)]}{\cosh[(i \pi -\alpha)/(2 \xi)]}\right]\label{Sm}.
\end{align}
\end{subequations}
The soliton scattering amplitudes $S_\iota(\alpha,\xi)$ are known \cite{ZZ79} to
become degenerate at the so-called reflectionless points 
\begin{equation}\label{rless}
\xi_n=1/n, \quad n=1,2\ldots
\end{equation}
 of the parameter $\xi$.
Specifically, for even $n=2 l$:
 \begin{align}
&S_+(\alpha,\xi_{2l})=-S_0(\alpha,\xi_{2l}), \quad S_-(\alpha,\xi_{2l})=S_0(\alpha,\xi_{2l}), \\\label{bos+}
&S_0(0,\xi_{2l})=S_-(0,\xi_{2l})=-1, \quad S_+(0,\xi_{2l})=1, 
\end{align}
while for  odd $n=2l-1$:
\begin{align}
&S_+(\alpha,\xi_{2l-1})=S_0(\alpha,\xi_{2l-1}), \quad S_-(\alpha,\xi_{2l-1})=-S_0(\alpha,\xi_{2l-1}), \\\label{bos-}
&S_0(0,\xi_{2l-1})=S_+(0,\xi_{2l-1})=-1, \quad S_-(0,\xi_{2l-1})=1,
\end{align}
with natural $l=1,2\ldots$. In particular, the two-particle scattering becomes trivial in the free-fermion point $\xi_1=1$: 
\begin{align}
S_+(\alpha,1)=S_0(\alpha,1)=-1,\quad S_-(\alpha,1)=1.
\end{align}

Perturbation of the two-folded sine-Gordon model  by the  term 
 $-2\Lambda \cos(\beta \varphi/2)$ with a small $\Lambda>0$ in the potential 
\eqref{V0} breaks the degeneracy between the vacua $|vac\rangle^{(a)}$, $a=0,1$.
The vacuum $|vac\rangle^{(1)}$ associated with potential minima at even $l$ in equation \eqref{Vm} decreases in energy 
and becomes the true ground state, while the vacuum $|vac\rangle^{(0)}$ associated with minima with odd $l$ increases in energy 
and transforms into the metastable false vacuum. 
 To the linear oder in $\Lambda\to +0$, the energy densities of the deformed vacua are given \cite{Baj2001,Roy2023} by the straightforward perturbative 
 formulas:
\begin{align}\label{vacd}
\mathcal{E}_{psG}^{(1)}(\Lambda)=\mathcal{E}_{sG}-2  \Lambda\,\mathcal{G}_{\beta/2}+O(\Lambda^2), \\
 \mathcal{E}_{psG}^{(0)}(\Lambda)=\mathcal{E}_{sG}+2 \Lambda\,\mathcal{G}_{\beta/2}+O(\Lambda^2),
\end{align}
 where $\mathcal{G}_\gamma$ is the vacuum expectation value of the exponential operator $e^{\pm i \gamma\varphi}$ 
 at $\Lambda=0$.
 The explicit expression for this expectation value  was found  by 
 Lukyanov and Zamolodchikov \cite{LZ97}:
 \begin{align}
\label{eq:G_gamma}
&{\cal G}_\gamma=
\bigg[\pi\mu\ \frac{\Gamma(1-\beta^2)}{\Gamma(\beta^2)}\bigg]^{\frac{\gamma^2}{1-\beta^2}}\\\nonumber
&\times
\exp\Bigg(\int_0^\infty\frac{d t}{t}\ \bigg(\frac{\sinh^2(2\gamma\beta t)}
{2\sinh(\beta^2 t) \sinh(t)\cosh\big((1-\beta^2) t\big)}-2\gamma^2\,e^{-2 t}   \Bigg)\ .
 \end{align}
 
 The margin  between the energies of  the false and true vacua induces the linear attractive potential $v(x_1,x_2)=\mathfrak{f}\cdot(x_2-x_1)$ between  two kinks located 
 near the points $x_1$ and $x_2$, such that  $x_2-x_1\gg m^{-1}$. Here  $\mathfrak{f}$ denotes the {\it string tension}:
 \begin{equation}
\mathfrak{f}= \mathcal{E}_{psG}^{(0)}(\Lambda)-\mathcal{E}_{psG}^{(1)}(\Lambda)=f+O(\Lambda^3),
 \end{equation}
 where 
  $
 f=4  \Lambda\,\mathcal{G}_{\beta/2}.$ 
   In what follows, we will widely use instead of  the parameter $f$ its dimensionless counterpart 
  \begin{equation}\label{la}
  \lambda=\frac{f}{m^2}.
  \end{equation}
  
  The linear attractive potential leads to confinement of kinks, which become coupled into the `meson' bound states 
  \cite{Mus_2004,FonZam2003,FZ06}.  The meson states $|\pi_{s,\iota,n}(P)\rangle$ can be classified by the momentum $P$,
  isospin $s=0,\pm 1$, parity $\iota=0,\pm$, and the quantum number $n=1,2,3\ldots$.  The quantum numbers $s$ and $\iota$
  are not independent: $\iota=0$ for $s=\pm1$, and $\iota=\pm$ at $s=0$.

  At $\Lambda=0$, the meson state 
  $|\pi_{s,\iota,n}(P)\rangle$ decouples into some linear combinations of the two-kink states:
  \begin{align}
 & |\pi_{s=1,\iota=0,n}(P)\rangle\to |K_{10}(\alpha_1)K_{01}(\alpha_2)\rangle_{1/2,1/2},\\
  & |\pi_{s=-1,\iota=0,n}(P)\rangle\to |K_{10}(\alpha_1)K_{01}(\alpha_2)\rangle_{-1/2,-1/2},\\
   & |\pi_{s=0,\iota=\pm,n}(P)\rangle\to |K_{10}(\alpha_1)K_{01}(\alpha_2)\rangle_{\pm},
  \end{align}
 with $p(\alpha_1)+p(\alpha_2)=P$.  So, the meson modes  with the isospin value $s=1$  ($s=-1$) represent  the bound states
 of two solitons (antisolitons), while the two meson modes with $s=0$  represent the soliton-antisoliton bound states with 
 different parities $\iota=\pm$.
 
 The mesons $|\pi_{s,\iota,n}(P)\rangle$ are  relativistic particles with the dispersion law:
 \begin{equation}
 E_{\iota,n}(P)=\sqrt{M_{\iota,n}^2+P^2},
 \end{equation}
 where $M_{\iota,n}$ are the meson masses. 
 
 Since the perturbed sine-Gordon model \eqref{psG0} is not integrable, the analytic calculation of the meson masses $M_{\iota,n}$ 
 is possible only by means of some perturbative technique. In this paper we  focus on the weak confinement regime
 of the perturbed sine-Gordon model \eqref{psG0}, \eqref{V0}, in which the constant $\Lambda>0$ is treated as a small parameter 
 of the theory. 
\section{Weak coupling expansions for the meson masses \label{Sec:WCE}} 
At first sight, it would be natural to use the FFPT for studying the  confinement phenomenon in the
non-integrable  deformations of integrable QFTs. It was realized, however, already in \cite{Del96,Del98} that the original version of the FFPT introduced in \cite{Del96} cannot be applied directly
to the confinement problem, since the confinement transition changes the particle content of the theory. In order to 
 circumvent this problem, two different perturbative techniques for calculation of the meson masses in the 
 weak confinement regime were developed. 
 
 The first more systematic and sophisticated technique exploits   the Bethe-Salpeter equation. The main advantage
 of this  approach  is that it allows one to account for the  long-range attraction between confined particles already in the 
 zeroth order of the perturbation theory. In  high-energy physics, the Bethe-Salpeter equation was applied to the confinement problem in the two-dimensional QCD with infinite number of colours by t'Hooft \cite{Hooft74}. Fonseca and Zamolodchikov \cite{FonZam2003,FZ06} derived the Bethe-Salpeter equation 
for the Ising field theory (IFT) and applied it to the calculation of the meson masses in this model in the limit of a weak magnetic field. 
 Later the technique based on the Bethe-Salpeter equation has been used for calculation of the meson energy spectra
 in three other (properly deformed) models exhibiting the kink confinement: in the quantum Ising spin chain \cite{Rut08a}, 
 in the gapped antiferromagnetic XXZ spin-1/2 chain \cite{Rut22}, and 
 in the three-state Potts field theory \cite{Rut23}.
 
It was  shown by means of this technique, 
 that the meson masses $M_n(\lambda)$ in the IFT in the weak confinement regime can be described by two asymptotic weak-coupling expansions in the 
 dimensionless small parameter $\lambda\to +0$, which is  defined by equation \eqref{la}.
 \begin{enumerate}
 \item
 The {\it low-energy expansion} in fractional powers of $\lambda$ describes the meson  masses 
 $M_n(\lambda)$ slightly above the edge point $2m$, i.e. at  $n\ll \lambda^{-1}$ and $\lambda\to+0$. 
 Many terms in this expansion were obtained by Fonseca and Zamolodchikov \cite{FZ06}. Two leading ones
 read:
   \begin{equation}\label{lE_Is}
  M_{n}(\lambda)= 2m +\lambda^{2/3}\, m \,z_n+O(\lambda^{4/3}),
  \end{equation}
  where $-z_n$ are the zeroes of the Airy function, $\mathrm{Ai}(-z_n)=0$. These two terms of the IFT low-energy
  expansion reproduce the old result of McCoy and Wu \cite{McCoy78}.
 \item The  {\it semiclassical expansion} \cite{FZ06,Rut05} in integer powers of $\lambda$ describes the masses of highly excited meson states  with $n\gg1$. It is convenient to parametrise the masses of  such mesons by the set of  rapidities $\alpha_n$:
 \begin{equation}\label{MiotIs}
 M_{n}(\lambda)=2 m \cosh \alpha_{n}.
 \end{equation}
 These rapidities must solve the transcendent equation with the power series in $\lambda$ on the right-hand side. 
 To the leading order in $\lambda$, this equation reads:
 \begin{equation}\label{semicl_Is}
\sinh(2 \alpha_{n})  -2 \alpha_{n}=\lambda[2\pi (n-1/4)]+O(\lambda^2).
  \end{equation} 
  \end{enumerate}
  
The second perturbative technique is not so powerful and rigorous, but, instead, rather heuristic and intuitive.  
Nevertheless, it allows one in a very simple way to recover the initial parts of  the weak coupling expansions
for the meson masses beyond the order $O(\lambda^2)$.
This technique can be viewed as a  generalisation of the phenomenological scenario of confinement, which was introduced by McCoy and Wu \cite{McCoy78} in 1978.

Proceeding to the description of this heuristic  technique, let us
consider a system of two relativistic particles of the mass $m$ moving in a line and attracting one another with a linear potential.
Their Hamiltonian reads:
\begin{equation}\label{H2}
 H(x_1,x_2,p_1,p_2)=\sqrt{p_1^2+m^2}+\sqrt{p_2^2+m^2}+f |x_1-x_2|,
\end{equation} 
where $x_1,x_2\in \mathbb{R}$, and $f>0$.
Since the total momentum of two particles $P=p_1+p_2$ is conserved, one can concentrate on their dynamics 
for the states with zero total momentum $P=0$. Then, the relative motion of two particles is determined by 
the reduced Hamiltonian:
 \begin{equation}\label{efH}
 \widetilde{H}(x,p)=2\sqrt{p^2+m^2}+f |x|,
\end{equation} 
where $x=x_1-x_2\in \mathbb{R}$. The variables $x$  and $p$ in \eqref{efH} represent the canonical coordinate and momentum.
The classical trajectories of the system \eqref{efH} are formed by periodically repeated cycles. Each cycle starts from the collision between two particles, and then follows by their subsequent movement  with a constant acceleration.

Let us now turn to the quantization of the simple classical dynamics determined by 
the Hamiltonian \eqref{efH}. The quantum version of the system \eqref{efH} has a discrete energy spectrum, and the energies of the two-particle bound states can be identified with the 
meson masses $M_n$, with  $n=1,2,\ldots$.  

For $n\gg 1$, the masses  can be obtained by means of the semiclassical 
quantization. Exploiting parametrization \eqref{MiotIs}, one 
obtains  in the straightforward fashion  \cite{FZ06,Rut05} the constraint on the rapidities $\alpha_n$ following from the Bohr-Sommerfeld quantization rule. 
The result depends on the particle statistics. If two particles are fermions, the Bohr-Sommerfeld quantization 
condition leads to the equation:
 \begin{equation}\label{semicl_IsF}
\sinh(2 \alpha_{n})  -2 \alpha_{n}=\lambda[2\pi (n-1/4)],
  \end{equation} 
  in agreement with \eqref{semicl_Is}. In the case of two bosons, one gets, instead:
  \begin{equation}\label{frB}
  \sinh(2 \alpha_{n})  -2 \alpha_{n}=\lambda[2\pi (n-3/4)].
  \end{equation}

A different quantization scheme should  be applied for calculation of the masses $M_n(\lambda)$ of light mesons,
such that
\[
\frac{M_n(\lambda)}{2 m}-1 \ll 1. 
\]
This strong inequality implies, that the momentum $p$ in the reduced Hamiltonian \eqref{efH} is small 
$|p|\ll m$, and, therefore,  one can replace this Hamiltonian by its non-relativistic approximation:
 \begin{equation}\label{efHb}
 \widetilde{H}(x,p)= 2m +\frac{p^2}{m}+f |x|+O(p^4).
\end{equation} 
After canonical quantization $p\to -i \partial_x$, the reduced Hamiltonian \eqref{efHb} transforms into the
Schr\"odinger Hamiltonian operator 
 \begin{equation}\label{Hop}
\hat{H}=2m - \frac{ \partial_x^2}{m} +f  |x|.
\end{equation} 
The set of its eigenvalues again depends on the particle statistics. If the two particles are fermions, the Hamiltonian
operator \eqref{Hop} acts in the space of odd wave functions, $\psi(-x)=-\psi(x)$,
and the meson masses are given by formula \eqref{lE_Is}. In the case of two bosons, the wave function must be even, $\psi(-x)=\psi(x)$,
and the meson masses are, instead, determined by the equation
  \begin{equation}\label{lE_IsB}
  M_{n}(\lambda)= 2m +\lambda^{2/3}\, m \,z_n'+O(\lambda^{4/3}),
  \end{equation}
  where $-z'_n$ denote the zeroes of the derivative of the Airy function, 
\[
\mathrm{Ai}'(z)\big|_{z=-z'_n}=0.
\]  

In summary, two asymptotic expansions describing the meson masses in the IFT in the weak confinement regime
can be derived by means of the systematic perturbative technique  based on the Bethe-Salpeter
equation. Few initial terms in these expansions can be recovered in a very simple way by means of the  quantization 
(semiclassical or canonical) of the classical dynamics of two particles, which is  determined by the Hamiltonian \eqref{H2}.

It is important to note, that the IFT has a very specific property: the kink topological excitations do not interact in the 
deconfined phase at zero magnetic field behaving like free fermions. In contrast, in many other integrable models, including the 
sine-Gordon field theory, elementary excitations strongly interact already in the deconfined phase at short distances, and this interaction  is encoded in the non-trivial scattering matrix. Fortunately, the outlined above heuristic perturbative technique, upon  minor modifications, can be also applied to such models. This modified heuristic technique 
was first introduced for the calculation of the meson masses in the  Potts field theory \cite{RutP09}, and 
later \cite{Rut18,Rut22} applied  to the calculation of the meson energy spectra in the antiferromagnetic 
XXZ spin chain perturbed by the staggered longitudinal magnetic field.  In paper \cite{Rut22}, this  technique 
was described in much detail, and   results for the 
meson energy spectra obtained by this method were confirmed  by  more rigorous calculations exploiting the perturbative solution of the Bethe-Salpeter equation.
In the rest of this section, we will discuss very briefly this modified heuristic technique. More details can be found in  \cite{Rut22}.

In integrable QFT, the short range interaction between particles is described by
the Faddeev-Zamolodchikov commutation relation, which contains  the elastic two-particle scattering matrix. 
Since the latter can be diagonalised by means of an appropriate unitary transform, it is sufficient 
to concentrate on the case of the one-channel scattering. 
In this case, the Faddeev-Zamolodchikov commutation relation reduces to the form:
\begin{equation}
Z^*(\alpha_1)Z^*(\alpha_2)=S(\alpha_1-\alpha_2) Z^*(\alpha_2)Z^*(\alpha_1),
\end{equation}
where $Z^*(\alpha)$ is the operator creating the particle with rapidity $\alpha$, and  $S(\alpha)$
is the scattering amplitude. It follows from the unitarity condition 
\begin{equation}\label{Un}
S(\alpha) S(-\alpha)=1,
\end{equation}
 that the scattering amplitude 
 can be written in the form:
\begin{equation}
S(\alpha)=- e^{i\theta(\alpha)},
\end{equation}
where  $\theta(\alpha)$ is the scattering phase. We will require, that this scattering phase is an odd function
of the rapidity: 
$\theta(-\alpha)=-\theta(\alpha)$.

The non-trivial two-particle scattering modifies the semiclassical quantization of the classical dynamics 
determined by the Hamiltonian \eqref{efH}: the  scattering phase of two colliding particles
 must be added to the left-hand side of the Bohr-Sommerfeld condition, as it was explained in \cite{RutP09,Rut22}.
 As the result, the semiclassical formulas \eqref{MiotIs}, \eqref{semicl_IsF} for the meson masses modify to the form
 (cf. eqs. (171), (172) in  \cite{Rut22}):
\begin{align}
&M_n(\lambda)=2 m \cosh \alpha_n,\\\label{secF}
&\sinh(2 \alpha_{n})  -2 \alpha_{n}=\lambda[2\pi (n-1/4)-\theta(2\alpha_n)],
\end{align}
with $n\gg1$.

It follows from \eqref{Un}, that $S(0)^2=1$. Accordingly,  one should distinguish the fermionic $S(0)=-1$, and 
bosonic $S(0)=1$ cases. In the fermionic case, 
the scattering phase $\theta(\alpha)$ is continuous at $\alpha\in \mathbb{R}$, and 
$\theta(0)=0$. In the bosonic case, the scattering phase
$\theta(\alpha)$
 must have a 
discontinuity at the origin,
and one can set:
$\lim_{\alpha\to\pm0} \theta(\alpha)= \pm \pi.$
Upon the described above  definition of the scattering phase $\theta(\alpha)$, formula \eqref{secF} holds in both fermionic and bosonic cases. 

In the bosonic case, it is convenient to introduce the 'bosonic' scattering phase:
\begin{equation}\label{bsc}
\tilde{\theta}(\alpha)= {\theta}(\alpha)-\pi\, \mathrm{sign}\, \alpha,
\end{equation}
which is odd in $\alpha$, and continuous at the origin: $\tilde{\theta}(0)=0$. In terms of the 
bosonic scattering phase
\eqref{bsc}, 
equation \eqref{secF} modifies the form:
\begin{equation}\label{secB}
\sinh(2 \alpha_{n})  -2 \alpha_{n}=\lambda[2\pi (n-3/4)-\tilde{\theta}(2\alpha_n)].
\end{equation}
In the case of free bosons $\tilde{\theta}(\alpha)\equiv 0$, the latter equation reduces to \eqref{frB}.

The short-range interaction between  particles in the deconfined phase effects also the 
low energy expansion for the meson masses in the weak  confinement regime.  
In order to account for  this effect in the frame of the heuristic approach outlined  above, one should add some 
short-range interaction  potential $u(x_1-x_2)$   in the classical phenomenological Hamiltonian \eqref{H2}.
After the canonical quantization, the potential  $u(x)$ emerges on the right-hand side of the Schr\"odinger Hamiltonian operator \eqref{Hop}. It was shown in \cite{Rut22} in the fermionic case $S(0)=-1$, that this leads to the 
additional  term $\sim \lambda$ in the  low-energy 
expansion \eqref{lE_Is} for the meson masses (see equation (187) in \cite{Rut22}): 
 \begin{equation}\label{lEF}
  M_{n}(\lambda)= 2m +\lambda^{2/3}\, m \,z_n+\lambda \,m^2 \,a+O(\lambda^{4/3}),
  \end{equation}
where ${a}$ denotes the scattering length \cite{taylor2012scattering}, which is defined as:
  \begin{equation}
    a=-m^{-1}\partial_\alpha \theta(\alpha)\Big|_{\alpha=+0}.
  \end{equation}
  The same   term $\sim\lambda$  appears as well in the low-energy expansion
  \eqref{lE_IsB}  in the bosonic case  $S(0)=1$:
   \begin{equation}\label{lEB}
  M_{n}(\lambda)= 2m +\lambda^{2/3}\, m \,z_n'+\lambda \,m^2 \,a+O(\lambda^{4/3}).
  \end{equation}
It is important to note, that  the derivation of formula \eqref{lEF} described in \cite{Rut22} was 
essentially based on the assumption,
that the scattering phase $\theta(\alpha)$ smoothly depends on $\alpha$ at  $|\alpha|\lesssim1$.

Now let us apply the general results described above to the dsG model \eqref{psG0}. 
In the {\it weak confinement regime} at $\Lambda\to+0$, the meson masses in the dsG model
admit two different asymptotic expansions
 in the dimensionless parameter $\lambda={ f}/{m^2}$. 
 \begin{enumerate}
\item
At large $n\gg1$ and small $\lambda\ll1$, the meson masses can be described 
at any $\xi>0$ by the {\it semiclassical} 
 asymptotic expansion in  integer powers of $\lambda$. The meson masses in this semiclassical regime 
 are determined by the formula
 \begin{equation}\label{Miot}
 M_{\iota,n}(\lambda,\xi)=2 m \cosh \alpha_{\iota,n},
 \end{equation}
in terms of the  rapidities  $\alpha_{\iota,n}$, which  solve the transcendent equation
 \begin{equation}\label{semicl}
\sinh(2 \alpha_{\iota,n})  -2 \alpha_{\iota,n}=\lambda[2\pi (n-1/4)-\theta_\iota(2 \alpha_{\iota,n},\xi)]+O(\lambda^2),
  \end{equation} 
  and the two-kink scattering phases $\theta_\iota(\alpha,\xi)$ are given by equations \eqref{th0}, \eqref{thet}.
  \item At a fixed $n\sim 1$ and $\lambda\to +0$ the {\it low-energy} asymptotic expansion in fractional powers 
  of $\lambda$ holds. Three  leading terms of this expansion read:
  \begin{equation}\label{lE}
  M_{\iota,n}(\lambda,\xi)= 2m +\lambda^{2/3}\, m \,z_n+\lambda \,m^2 \,a_\iota(\xi)+O(\lambda^{4/3}),
  \end{equation}
  where $-z_n$ are the zeroes of the Airy function, $\mathrm{Ai}(-z_n)=0$, and $a_\iota(\xi)$ denotes 
  the scattering length:
  \begin{equation}\label{scl}
  a_\iota(\xi)=-m^{-1}\partial_\alpha \theta_\iota(\alpha,\xi)\Big|_{\alpha=0}.
  \end{equation}
 In the explicit form, the scattering lengths $ a_\iota(\xi)$ in the sine-Gordon model read:
  \begin{align}\label{a0}
   & a_0(\xi)=-\frac{1}{2\pi  m \xi}\int_{-\infty}^\infty dx \, \left[
    1-\tanh\frac{x}{2\xi}\, \coth\frac{x}{2}
    \right],\\\label{ap}
    & a_+(\xi)=a_0(\xi)+\frac{1}{m \xi}\, \cot\frac{\pi}{2\xi},\\\label{am}
    &  a_-(\xi)=a_0(\xi)-\frac{1}{m \xi}\, \tan\frac{\pi}{2\xi}.
  \end{align}
  \end{enumerate}
    
 In the deformed sine-Gordon model, the low-energy expansion  \eqref{lE} 
 for the mesons with parity $\iota=0$ holds at any $\xi>0$. This asymptotical formula 
 describes also the masses of light mesons with parities $\iota=\pm$ for the values of the parameter  
 $\xi$, which are not too close to the 
reflectionless points \eqref{rless}.   Indeed, the soliton-antisoliton scattering lengths $a_+(\xi)$ 
  and  $a_-(\xi)$ given by formulas \eqref{ap}, and \eqref{am}  diverge, respectively,  at $\xi^{-1}=2l$, and at $\xi^{-1}=2l-1$, with  natural $l$.
  Directly at the reflectionless points, formula \eqref{lE}  must be  modified in the following way. 
  
At $\xi=\frac{1}{2l}$, the soliton-antisoliton scattering in the $\iota=+$ channel becomes bosonic, $S_+(0,\xi_{2l})=1$, see 
equation \eqref{bos+}. Therefore, one should use the bosonic low-energy expansion \eqref{lEB} for the mesons 
with parity $\iota=+$  in this case. 
For the bosonic scattering phase \eqref{bsc}, we get from \eqref{Sp}, \eqref{Spp}:
  \begin{equation}\label{tht}
\tilde{\theta}_+(\alpha,\xi_{2l})={\theta}_0(\alpha,\xi_{2l}).
  \end{equation}
Thus, the low-energy expansion for the meson masses  $M_{+,n}(\lambda,\xi)$ at $\xi=\xi_{2l}$ takes the form:
   \begin{equation}\label{lEa1}
  M_{+,n}(\lambda,\xi_{2l})= 2m +\lambda^{2/3}\, m \,z'_n+\lambda \,m^2\, a_0(\xi_{2l})+O(\lambda^{4/3}),
  \end{equation} 
  where $-z'_n$ denotes the zeroes of the derivative of the Airy function.

In turn, the transcendent equation  \eqref{semicl}, which determines the 
semiclassical expansion for the   masses $ M_{+,n}(\lambda,\xi_{2l})$, can be modified 
due to  \eqref{secB}, \eqref{tht} to the  more convenient form:
  \begin{equation}\label{sem2}
  \sinh(2 \alpha_{+,n})  -2 \alpha_{+,n}=\lambda[2\pi (n-3/4)-\theta_0(2 \alpha_{+,n},\xi_{2l})]+O(\lambda^2).
  \end{equation}

  At $\xi=\xi_{2l-1}$, with $l=1,2,\ldots$, the  soliton-antisoliton scattering in the $\iota=-$ channel becomes bosonic 
  due to \eqref{bos-}, and $\tilde{\theta}_-(\alpha,\xi_{2l-1})={\theta}_0(\alpha,\xi_{2l-1})$.
 Accordingly, the  
 low-energy expansion for the masses   $M_{-,n}(\lambda,\xi_{2l-1})$ modifies in the 
similar way:
   \begin{equation}\label{lEa}
  M_{-,n}(\lambda,\xi_{2l-1})= 2m +\lambda^{2/3}\, m \,z'_n+\lambda\, m^2\, a_0(\xi_{2l-1})+O(\lambda^{4/3}).
  \end{equation} 
  In turn, the transcendent equation  \eqref{semicl} reduces at $\xi=\xi_{2l-1}$ and $\iota=-$  to the form:
    \begin{equation}
  \sinh(2 \alpha_{-,n})  -2 \alpha_{-,n}=\lambda[2\pi (n-3/4)-\theta_0(2 \alpha_{-,n},\xi_{2l-1})]+O(\lambda^2).
  \end{equation}

  \section{Low-energy expansions at $\xi^{-1}$ close to natural numbers \label{Sec_LEE}}
    At $\xi^{-1}$ close to the natural numbers, the low energy expansion \eqref{lE} for the meson masses $  M_{\pm,n}(\lambda,\xi)$ must be further modified. 
  In this section, we will obtain the modified low-energy expansion for the meson masses $ M_{-,n}(\lambda,\xi)$, which
   will replace \eqref{lE} (for $\iota=-$) 
  in the crossover regions of the parameter $\xi$ close to $\xi_{2l-1}$, i.e. for 
  \begin{equation}\label{xiod}
  \xi=\frac{1}{2l-1}+\delta \xi, \text{ with }|\delta \xi|\ll 1,
   \end{equation} 
   where $l=1,2\ldots$.  The similar modified low-energy  expansion for the meson masses $M_{+,n}(\lambda,\xi)$ at   $\xi=\frac{1}{2l}+\delta\xi$, with natural $l$ and small $|\delta\xi|$, will be presented by the end of the section.

   At $\xi=\xi_{2l-1}$, the two-kink scattering phases $\theta_+(\alpha,\xi)$ and  $\theta_0(\alpha,\xi)$  coincide:   
     \begin{equation} 
  \theta_+ \left(\alpha,\xi_{2l-1}\right)=\theta_0(\alpha,\xi_{2l-1}).
      \end{equation}  
 This odd function smoothly depends  on the rapidity $\alpha$, and at small $|\alpha|\ll1$ behaves as:    
     \begin{equation} \label{S0a}
\theta_0(\alpha,\xi_{2l-1})= \alpha\, \partial_\alpha\theta_0(\alpha,\xi_{2l-1})|_{\alpha=0}  +O(\alpha^3).
  \end{equation}      
Tuning the parameter $\xi$  from the point $\xi_{2l-1}$  by a small $\delta\xi$ 
induces only  the uniformly small in $\alpha$ variations  $\delta \theta_0(\alpha,\xi)$ of this scattering phase: $|\delta \theta_0(\alpha,\xi)|<C|\delta \xi|$, where the positive number $C$ does not depend on $\alpha$.

\begin{figure}[htb]
\includegraphics[width=\linewidth, angle=00]{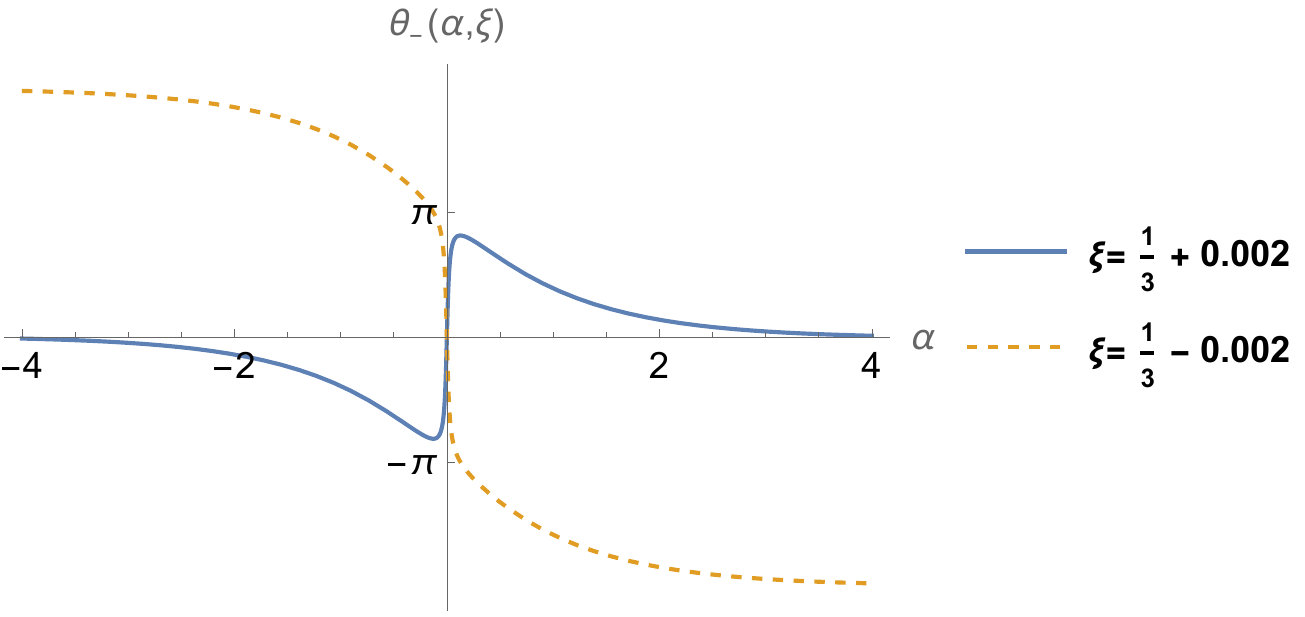}

\caption{Plot of the scattering phase $\theta_-(\alpha,\xi)$ determined by 
\eqref{Sp} at $\xi=\frac{1}{3}+ 0.002,$ and at $\xi=\frac{1}{3}-0.002.$
\label{fig:Thetm} } 
\end{figure}

   In contrast, the scattering phase $\theta_-(\alpha,\xi)$ at $\xi$ in the crossover region  \eqref{xiod}  displays a fast variation at small real $\alpha$. For two values of $\xi$ close to $1/3$, this fast variation of the scattering phase $\theta_-(\alpha,\xi)$ 
   near the origin is illustrated in 
   Figure \ref{fig:Thetm}. 
   
   It was noticed in Section \ref{Sec:WCE}, that a smooth variation of the 
   scattering phase $\theta(\alpha)$ at small $\alpha$ was the essential requirement exploited in the derivation of the low-energy expansion \eqref{lEF}. Since this requirement is broken for the scattering phase $\theta_-(\alpha,\xi)$ at $\xi$ in the crossover region  \eqref{xiod}, the low-energy expansion \eqref{lE} for the masses $M_{-,n}(\lambda,\xi)$ of light
mesons cannot be used there. We proceed now to the derivation of the different low-energy asymptotic formula for the meson masses $M_{-,n}(\lambda,\xi)$, which will replace   \eqref{lE} in the crossover regions \eqref{xiod}.

First, let us note, that the 
 fast variation  of the scattering phase $\theta_-(\alpha,\xi)$ at $\xi$ close to $\xi_{2l-1}$ 
 is induced, mathematically,  by approaching to the real $\alpha$-axis of the  pole and zero of the scattering amplitude   $S_-(\alpha,\xi)$, which lie on the imaginary $\alpha$-axis  at the points $\alpha_B=- i \pi (2l-1)\delta \xi$, and  
 $-\alpha_B$, respectively.  At $\delta \xi<0$, the pole $\alpha_B$ enters the physical sheet 
 $0<\mathrm{Im}\, \alpha<\pi$, indicating  appearance  of the $(2l-1)$th breather in the theory at $\xi<\xi_{2l-1}$.

Taking into account the above mentioned analytical properties of the the  scattering amplitude  $S_-(\alpha,\xi)$, we rewrite 
the latter  for $\xi$ in the vicinity of the point $\xi_{2l-1}$ in the form
        \begin{equation}\label{Scmin}
    S_-(\alpha,\xi_{2l-1}+\delta \xi)=-S_-^{(reg)}(\alpha,\xi_{2l-1}+\delta \xi)\, \frac{\alpha-i \pi(2l-1)\,\delta\xi}{\alpha+i \pi(2l-1)\,\delta\xi}.
    \end{equation}  
The function $S_-^{(reg)}(\alpha,\xi_{2l-1}+\delta \xi)$ on the right-hand side  has at $|\delta \xi |\ll \xi_{2l-1}-\xi_{2l+1}$
 neither poles nor zeroes in  the narrow strip 
$ |\mathrm{Im}\,\alpha|<\gamma_l$ with $\gamma_l\approx {2\pi \xi_{2l-1}}$, and 
     \begin{equation}
     S_-^{(reg)}(0,\xi_{2l-1}+\delta \xi)=-1.
     \end{equation}   
       
   At small $\alpha$, 
    the kink  rapidity   becomes proportional to its momentum:
        \begin{equation}
   \alpha(p)=\frac{p}{m}+O(p^3),
       \end{equation}
  and       the  kink relativistic dispersion law \eqref{DL} can be approximated by its non-relativistic asymptotics:
     \begin{equation}
     \omega (p)=m+\frac{p^2}{2m}+O(p^4).
    \end{equation}   

As it was shown by McCoy and Wu \cite{McCoy78}, the low-energy expansions for the meson masses in relativistic QFTs   in the weak confinement regime can be effectively studied
in the frame of the non-relativistic quantum mechanics.
Following this strategy, let us first consider  a system of two non-relativistic  $\delta$-interacting bosons moving in a line. The Hamiltonian of the system reads:
       \begin{equation}\label{Ham2} 
  \mathcal{H}_0= 2 m-\frac{\partial_{x_1}^2}{2 m} -\frac{\partial_{x_2}^2}{2 m} +
c\,\delta(x_1-x_2),       
       \end{equation}
 where $c$ is the interaction constant.

The quantum state of two bosons can  be described by the 
wave-function $\Psi(x_1,x_2)$, with $x_1,x_2\in \mathbb{R}$, which is symmetric under permutation of the spatial coordinates 
$x_1,x_2$.  Up to the constant term $2m$, the Hamiltonian \eqref{Ham2}   coincides with the Hamiltonian of the Lieb-Liniger model in the two-particle
sector,  which is discussed in Appendix \ref{LibLin}.

The  solution of the Hamiltonian eigenvalue problem 
\begin{equation}
\mathcal{H}_0\,\Psi_{p_1,p_2}(x_1,x_1)={E}(p_1,p_2)\, \Psi_{p_1,p_2}(x_1,x_1)
\end{equation}
in the half-plane $-\infty<x_1<x_2<+\infty$ is given by the Bethe wave function:
\begin{equation}\label{Beig0}
\Psi_{p_1,p_2}(x_1,x_1)=\exp[i(p_1 x_1+p_2 x_2)]+\mathcal{S}(p_1-p_2) \exp[i(p_1 x_2+p_2 x_1)],
\end{equation}
where 
\begin{equation}\label{Sc2}
\mathcal{S}(p_1-p_2)=\frac{p_1-p_2 - i c \, m}{p_1-p_2 + i c \, m}
\end{equation}
is the two-particle scattering amplitude, and 
\begin{equation}
{E}(p_1,p_2)=2 m+\frac{p_1^2}{2m}+\frac{p_1^2}{2m}
\end{equation}
is the energy of the state \eqref{Beig0}.

Upon the choice
 \begin{equation} \label{cxi}
c=(2l-1)\pi \,\delta\xi,
\end{equation}
 the scattering amplitude \eqref{Sc2}   coincides with the leading term of the 
 small-$\alpha$ asymptotics  of the scattering amplitude $S_-(\alpha_1-\alpha_2,\xi_{2l-1}+\delta \xi)$ of the sine-Gordon model
 \begin{equation}
 \mathcal{S}(p_1-p_2)=S_-(\alpha_1-\alpha_2,\xi_{2l-1}+\delta \xi)\Big|_{\alpha_{1,2}=p_{1,2}/m}+O(p_1-p_2).
    \end{equation}  
       
Let us  now  deform the Hamiltonian \eqref{Ham2} by adding the linear attractive potential acting between bosons:
\begin{equation}\label{Ham3} 
  \mathcal{H}(f)= 2 m-\frac{\partial_{x_1}^2}{2 m} -\frac{\partial_{x_2}^2}{2 m} +
c\,\delta(x_1-x_2)+f \,|x_1-x_2|, 
\end{equation}
where $f>0$ is the string tension. 
The energy spectrum  $\{{E}_n(f,c)\}_{n=1}^\infty$ of the particle bound states with zero total  momentum  can be easily found, as it is described in 
Appendix \ref{LibLin}. The final result reads:
\begin{equation}\label{Enfc}
E_n=2m+m \lambda^{2/3} t_n,
\end{equation}
where $\lambda=f/m^2$, and the numbers $t_n$ are the successive solutions of the transcendent equation 
\begin{equation}\label{trAi}
Y_\rho(t_n)=0, 
\end{equation}
with $Y_\rho(t)=\mathrm{Ai}'(-t)-\rho \,\mathrm{Ai}(-t)$.
Here $\mathrm{Ai}(z)$ is the Airy function,
and
\begin{equation}\label{rhA}
\rho=\frac{c}{2 \lambda^{1/3}}.
\end{equation}

Formulas \eqref{Enfc}-\eqref{rhA}, and \eqref{cxi} determine two leading terms in the modified low-energy expansion for the masses of mesons having
 the negative parity $\iota=-$ in the crossover region  \eqref{xiod}. 
One more term $\sim \lambda$ in this low-energy expansion can be determined following the lines described in 
\cite{Rut22}
by taking into account the smooth dependence of the function
 $S_-^{(reg)}(\alpha,\xi)$ on the rapidity $\alpha$ at $\alpha\to0$.  
Then, the modified low-energy expansion representing
in the crossover region \eqref{xiod}  the masses of mesons with parity $\iota=-$
 takes the form:
  \begin{equation}\label{Mmin}
 M_{-,n}(\lambda,\xi_{2l-1}+\delta\xi)=2m +m \lambda^{2/3} t_n +\lambda\, m^2\, 
 [a_0(\xi_{2l-1})+O(\delta \xi)]+O(\lambda^{4/3}),
 \end{equation}
 where  the scattering length $a_0(\xi)$  is given by \eqref{a0}.

Let us describe the meson masses predicted by equation \eqref{Mmin} in three limit cases.
 \begin{enumerate}
 \item At $\rho \gg 1$, the solutions of the secular equation \eqref{trAi} have the following asymptotics 
 \[
 t_n(\rho)=z_n- \rho^{-1}+O(\rho^{-3}).
 \]
 Accordingly, formula \eqref{Mmin} reduces 
 at  $\lambda^{1/3}\ll \xi-\xi_{2l-1}\ll1$  to the form
  \begin{align*}
 &M_{-,n}(\lambda,\xi)\\
 &=2m +m \lambda^{2/3} \left(z_n -\frac{2 \lambda^{1/3}}{c}+\ldots\right)+\lambda\, m^2\, a_0(\xi_{2l-1})+O(\lambda^{4/3})\\
 &=2m +m \lambda^{2/3} z_n +\lambda\, m^2\, \left[-\frac{2}{c\,m}+a_0(\xi)+\ldots\right]+O(\lambda^{4/3})\\
 &=2m +m \lambda^{2/3} z_n +\lambda\, m^2\, a_-(\xi)+O(\lambda^{4/3}),
 \end{align*}
 in agreement with \eqref{lE}.
 \item At $\rho=0$, solutions of equation \eqref{trAi} taken with the `minus' sign are just the zeroes of the derivative of the Airy function: $t_n(0)=z_n'$, $\mathrm{Ai}'(-z_n')=0$,
 and formula \eqref{Mmin} reduces to \eqref{lEa}.
 \item At a large negative $\rho$, the first solution $t_1(\rho)$ of equation \eqref{trAi} becomes negative and well separated from all other solutions
 $t_2(\rho),t_3(\rho),\ldots$.
 The following asymptotic formulas hold at $\rho\to -\infty$:
\begin{equation}
 t_1(\rho)=-\rho^2-\frac{1}{2\rho}+O(\rho^{-4}), 
 \end{equation}
 and for $n=2,3\ldots$: 
 \[t_n(\rho)=z_{n-1}-\rho^{-1}+O(\rho^{-3}).
 \]
 As the result, we obtain at $\lambda^{1/3}\ll -\delta\xi\ll1$ for  the  meson masses \eqref{Mmin}:
 \begin{align}\label{Mm1}
 &M_{-,1}(\lambda,\xi_{2l-1}+\delta\xi)=2m -\frac{m c^2}{4}\\\nonumber
 &+\lambda m\left[-\, \frac{1}{c}+m\,a_0(\xi_{2l-1})+O(\delta \xi)\right]+ \ldots\\\nonumber
 &=2m -\frac{\pi^2 m }{4}\left(\frac{\delta \xi}{\xi_{2l-1}}\right)^2+\lambda\, m \,\left[m\, a_0(\xi_{2l-1})
 -\frac{\xi_{2l-1}}{\pi  \delta \xi}+O(\delta\xi)\right]+ \ldots
 \end{align}
and 
 \begin{equation}
 M_{-,n}(\lambda,\xi_{2l-1}+\delta\xi)=2m +m \lambda^{2/3} z_{n-1} +\lambda\, m^2\, a_-(\xi_{2l-1}+\delta\xi)+O(\lambda^{4/3}),
 \end{equation}
 for $n=2,3,\ldots$.   At $l=1$, formula \eqref{Mm1} reduces to the form:
  \begin{equation}\label{M1mass}
 M_{-,1}(\lambda,\xi)\Big|_{\xi=1+\delta\xi}=2m -\frac{\pi^2 m }{4}\,\delta\xi^2
 +\lambda\, m \left[-\frac{1}{\pi  \delta \xi}+O(\delta\xi)\right]+ \ldots
  \end{equation}
 
 The lightest meson with the mass given by \eqref{Mm1} must be identified with the $j$th breather, with $j=2l-1$, perturbed by 
 the  term 
 $-2\Lambda \cos(\beta \varphi/2)$ in the potential  \eqref{V0}. 
Indeed, the mass $ m_{2l-1}^{(b)}(\xi)$  of this breather at $f=0$ and $\xi<\xi_{2l-1}$ is given by equation \eqref{mbr}:
 \[
 m_{2l-1}^{(b)}(\xi)=2 m \sin\left(\frac{\pi \xi}{2\,\xi_{2l-1}}\right). 
 \]
Two initial terms in  its  Taylor expansion at $\xi=\xi_{2l-1}$ read: 
 \begin{equation}\label{mb2}
  m_{2l-1}^{(b)}(\xi)=2m -\frac{\pi^2 m }{4}\left(\frac{ \xi-\xi_{2l-1}}{\xi_{2l-1}}\right)^2+O\big((\xi-\xi_{2l-1})^4\big).
 \end{equation}
 At $\lambda=0$, the last line of \eqref{Mm1} reduces to the right-hand side of \eqref{mb2}, in confirmation of the above statement.
 The linear in $\lambda$ term in the last line of \eqref{Mm1} represents the linear perturbation of the breather mass by the deforming potential  $-2\Lambda \cos(\beta \varphi/2)$.
 \end{enumerate}

 The linear in $\lambda$ correction to the breather masses can be independently calculated by means of the 
 FFPT. For the lightest breather, this $O(\lambda)$ correction was explicitly found 
 by Bajnok {\it {et al.}} \cite{Baj2001}, and recently represented in
 a compact form by Roy and Lukyanov \cite{Roy2023}. Combining  formulas (II.4), (II.5) in the Supplementary information 
 to Ref. \cite{Roy2023} with \eqref{mbr}, and \eqref{vacd}-\eqref{la}, yields:
 \begin{equation}\label{br1M}
   m_{1}^{(b)}(\xi,\lambda)=2 m \sin(\pi \xi/2)+\frac{\lambda\, m}{2\cos(\pi \xi/2)}+O(\lambda^2).
  \end{equation}
This formula represents  two initial terms of  the FFPT  expansion for the mass of the lightest breather in the dsG model 
\eqref{psG0}. It  holds in the interval  $\xi\in(0,1)$ for $\xi$ not too close to the free-fermion point $\xi_1=1$,
at which the second term  in the right-hand side  has a simple pole. The residue $(- \lambda\, m/\pi)$ 
at this pole is reproduced by our formula \eqref{M1mass}.

 The three initial terms of the modified low-energy expansion for the masses of the mesons with $\iota=+$ at $\xi$ close to $(2l)^{-1}$
 can be obtained by means of the same procedure. The only difference is that formula \eqref{cxi} must be now replaces by:
  \begin{equation*}
c=2l\,\pi \,\delta\xi.
\end{equation*}
 The final result reads:
   \begin{equation}\label{Msp}
 M_{+,n}(\lambda,\xi)=2m +m \lambda^{2/3} t_n +\lambda\, m^2\, [a_0(\xi_{2l})+O(\delta\xi)]+O(\lambda^{4/3}),
 \end{equation}
 where $\xi=\xi_{2l}+\delta\xi$,  and $|\delta\xi|\ll 1$.
 
 It follows from \eqref{Msp}, that at 
    \begin{equation}\label{xi2l}
 \lambda^{1/3}\ll \xi_{2l}-\xi\ll1,
  \end{equation}
  the masses of  mesons with $\iota=+$ behave as:
 \begin{equation}\label{Mp1}
 M_{+,1}(\lambda,\xi_{2l}+\delta\xi)=2m -\frac{\pi^2 m }{4}\left(\frac{\delta \xi}{\xi_{2l}}\right)^2+\lambda\, m \,
 \left[m\,a_0(\xi_{2l})-\frac{\xi_{2l}}{\pi  \delta \xi}+O(\delta\xi)\right]+ \ldots,
 \end{equation}
and 
 \begin{equation}
 M_{+,n}(\lambda,\xi_{2l}+\delta\xi)=2m +m \lambda^{2/3} z_{n-1} +\lambda\, m^2\, a_+(\xi_{2l}+\delta\xi)+O(\lambda^{4/3}),
 \end{equation}
 for $n=2,3,\ldots$. At $\xi$ satisfying \eqref{xi2l}, the lightest meson in this set is just the perturbed $(2l)$th breather. 
 
 So, upon decrease of the parameter $\xi$ in the crossover region close to the point $\xi_{2l-1}=1/(2l-1)$, the lightest meson with 
 the negative parity $\iota=-$ transforms into the $(2l-1)$th breather, which also has the negative parity according to \eqref{brs}.
 In turn, the lightest meson with the positive parity $\iota=+$ transforms upon decrease of $\xi$ close to the point $\xi_{2l}=1/(2l)$ 
 into the $(2l)$th breather, which parity is  positive as well. 
\section{Evolution of  meson masses with parameter  $\xi$\label{Sec:Ev}}
In this section we illustrate and compare the predictions for the meson masses given by different asymptotic expansions, 
which were obtained in the previous section. 
The emphasis will be made on studying of the variation of these masses upon tuning the 
parameter $\xi$.

\begin{figure}{}
\centering
\subfloat[ 
]
{
\includegraphics[width=.52\linewidth]{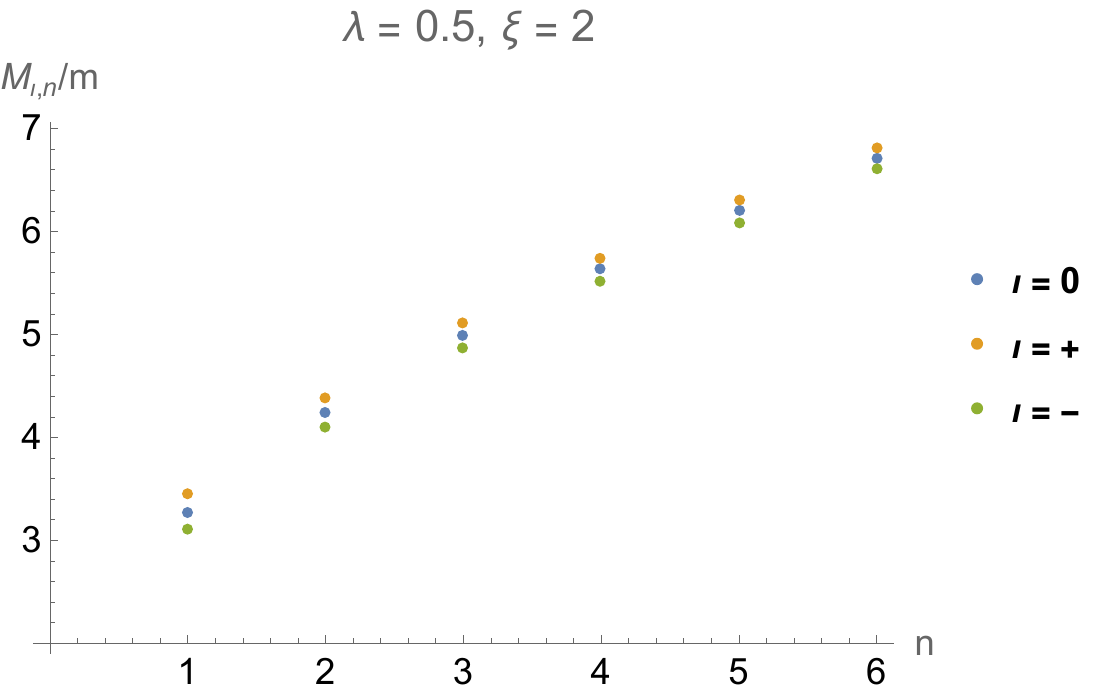}\label{fig:subfig0}}

\begin{minipage}{.48\textwidth}%
\subfloat[Subfigure 2 list of figures text][]{
\includegraphics[width=1.1\textwidth]{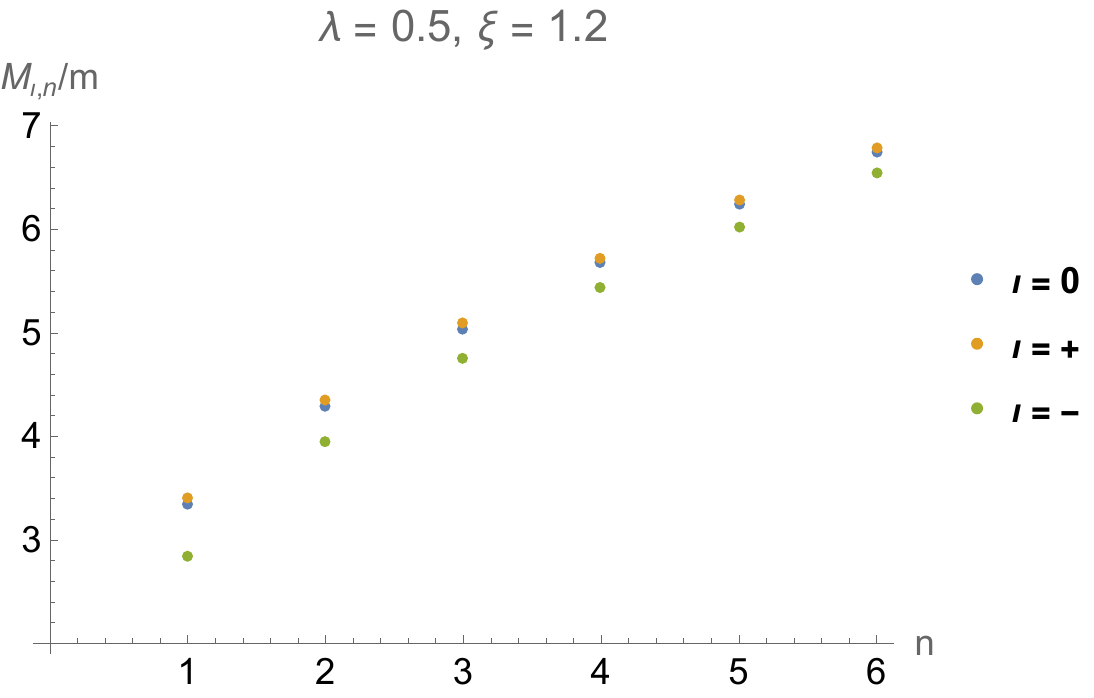}
\label{fig:subfig1}} \\
\subfloat[Subfigure 3 list of figures text][]{
\includegraphics[width=1.1\textwidth]{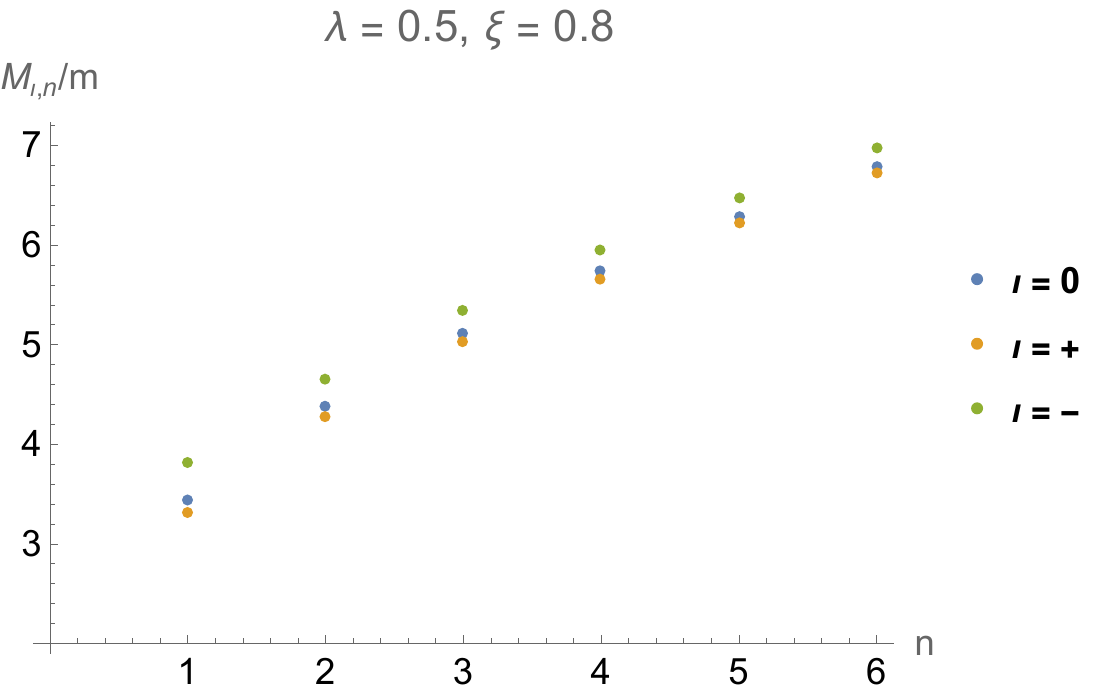}
\label{fig:subfig2}}
\end{minipage}
\qquad
\begin{minipage}{.46\textwidth}%
\subfloat[Subfigure 2 list of figures text][]{
\includegraphics[width=1.1\textwidth]{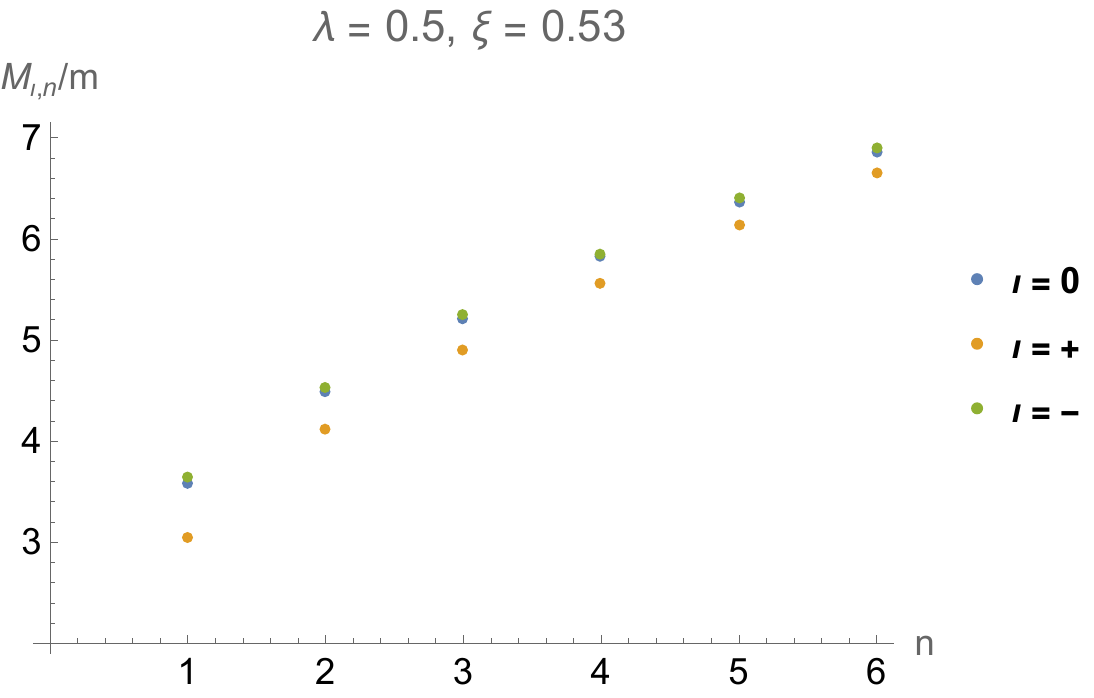}
\label{fig:subfig3}} \\
\subfloat[Subfigure 3 list of figures text][]{
\includegraphics[width=1.1\textwidth]{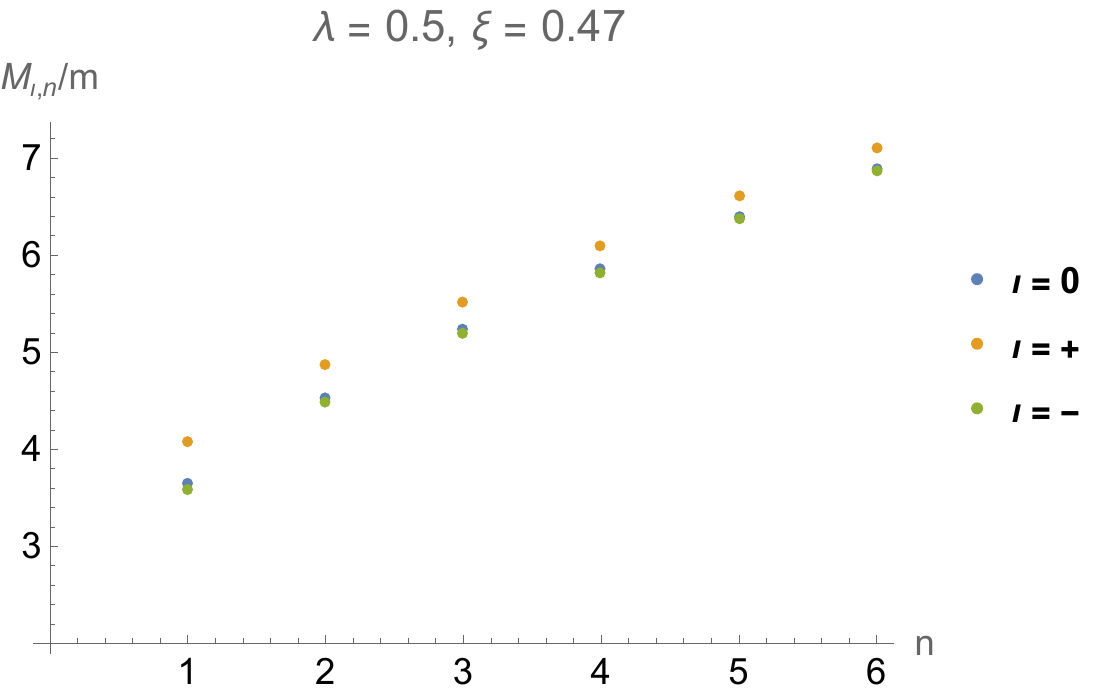}
\label{fig:subfig4}}
\end{minipage}%
\caption{Masses of mesons   $M_{\iota,n}(\lambda,\xi)$ calculated by 
semiclassical  formulae \eqref{Miot}, \eqref{semicl} at $\lambda=0.5$, $\iota=0,\pm$, and $n=1,\ldots, 6$
for (a) $\xi=2$, (b) $\xi=1.2$, (c) $\xi=0.8$, (d)  $\xi=0.53$, (e)  $\xi=0.47$.
\label{fig:Masses1}}
\end{figure}

Figure \ref{fig:Masses1} displays the masses $M_{\iota,n}(\lambda,\xi)$ of six lightest mesons determined by the  semiclassical 
formulas \eqref{Miot}, \eqref{semicl}
at the fixed value of the parameter $\lambda=0.5$ and five different values of the parameter $\xi$.
Note, that the semiclassical approximation a priori is supposed to work well only for the  mesons with large $n\gg1$,
while the masses of the lightest mesons should be instead described by the low-energy expansions.
It will be shown later, however, that in many cases, the predictions for the masses of the lightest mesons 
(with small $n=1,2,\ldots$) given by the semiclassical and low-energy expansions are numerically very close
to each other.\begin{footnote}{The high efficiency of the semiclassical expansions for prediction the masses of lightest mesons 
in several models exhibiting confinement were noticed in
papers \cite{Tak14,Kor16,Lagnese_2020,Mus22}, in which the meson masses were studied by direct numerical methods.}
\end{footnote}

The meson masses shown in Figure \ref{fig:Masses1} form triplets, in which the mesons are distinguished by 
the parity index $\iota=0,\pm$.
The
splitting of the meson masses inside each triplet is caused by the $\iota$-dependent scattering phase
$\theta_\iota(2 \alpha_{\iota,n},\xi)$ on the right-hand sides of \eqref{semicl}.

At  $\xi>1$, the meson masses in each triplet are ordered in the following way:
 \begin{equation}\label{od1}
M_{-,n}(\lambda,\xi)<M_{0,n}(\lambda,\xi)<M_{+,n}(\lambda,\xi). 
\end{equation}
This situation is illustrated in Figures \ref{fig:subfig0}, and \ref{fig:subfig1}, which correspond to the  $\xi$-parameter values 
$\xi=2$, and $\xi=1.2$, respectively.

Directly at the free-fermion point $\xi=1$, two scattering phases vanish: 
\[
\lim_{\xi\to1}\theta_\iota(\alpha,\xi)=0, \quad \text{ for }\iota=0,+,
\]
while the third one reduces to the step function:
 \begin{equation}
\lim_{\xi\to1\pm 0}\theta_-(\alpha,\xi)=\pm \pi \, \mathrm{sign}\, \alpha.
\end{equation}
Accordingly, the masses of the mesons with parities $\iota=0,+$, continuously depend on $\xi$ near the free-fermion 
point $\xi=1$, and $M_{0,n}(\lambda,1)=M_{+,n}(\lambda,1)$. In contrast, the masses of the mesons with 
negative parity  display a discontinuity at $\xi=1$, such that:
 \begin{equation}
 \lim_{\xi\to1- 0} M_{-,n}(\lambda,\xi)= \lim_{\xi\to1+ 0} M_{-,n+1}(\lambda,\xi).
\end{equation}
As the result, the masses of the mesons at $\xi\in(1/2,1)$ in each triplet are ordered due to
 \begin{equation}\label{od2}
M_{+,n}(\lambda,\xi)<M_{0,n}(\lambda,\xi)<M_{-,n}(\lambda,\xi),  
\end{equation}
as one can see in Figures \ref{fig:subfig2}, and \ref{fig:subfig3}.

The next reordering of the meson masses upon decrease of the parameter $\xi$ takes place at $\xi=1/2$, 
and the order \eqref{od1} is restored in the interval $\xi\in (1/3,1/2)$, as it is shown in Figure \ref{fig:subfig4}.
Directly at $\xi=1/2$ one gets from  \eqref{Miot}, \eqref{semicl}, and  \eqref{thet}:  $M_{0,n}(\lambda,1/2)=M_{-,n}(\lambda,1/2)$, and 
 \begin{equation}
 \lim_{\xi\to1/2- 0} M_{+,n}(\lambda,\xi)= \lim_{\xi\to1/2+ 0} M_{+,n+1}(\lambda,\xi).
\end{equation}

In general, the meson masses in each triplet are ordered according to \eqref{od1} at $\xi\in\left((2l+1)^{-1}, (2l)^{-1}\right)$, and 
due to \eqref{od2} at $\xi\in\left((2l)^{-1}, (2l-1)^{-1}\right)$, with natural $l$. 

\begin{figure}
\centering
\subfloat[ 
]
{
\includegraphics[width=.7\linewidth]{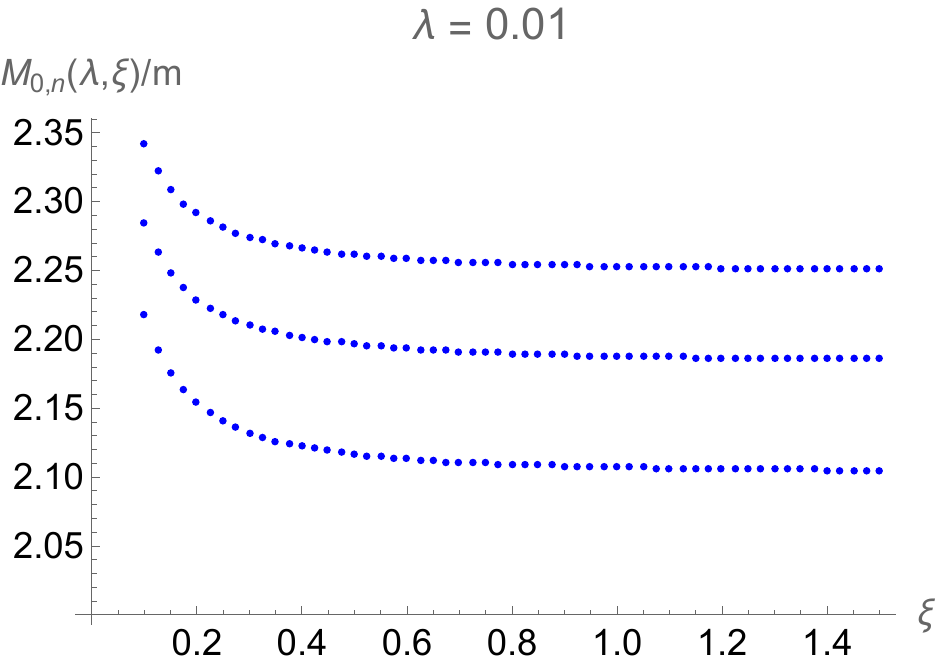}\label{fig:M0}}

\subfloat[
]
	{
\includegraphics[width=.7\linewidth]{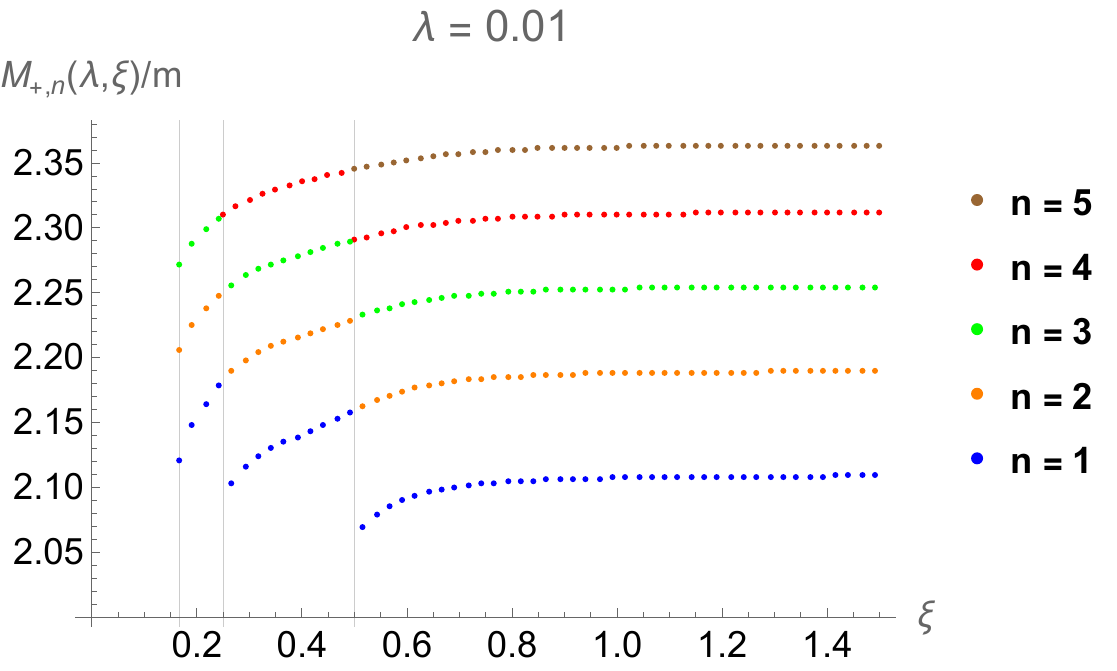}\label{fig:M1}}

\subfloat[
]
	{
\includegraphics[width=.7\linewidth]{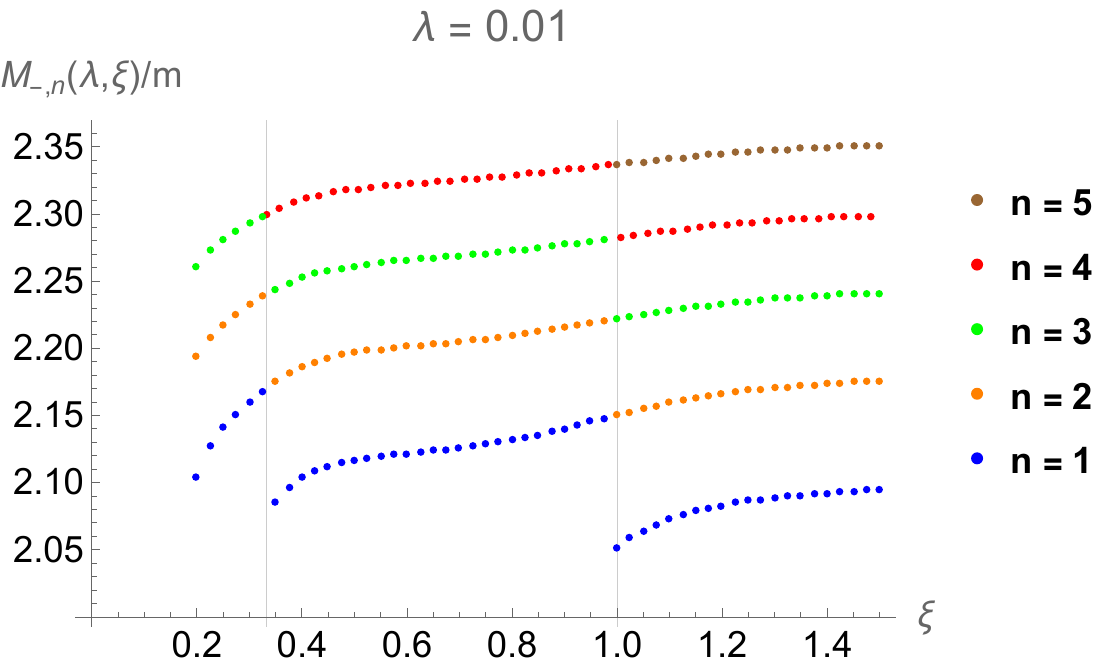} \label{fig:M2}}

\caption{Variation with $\xi$ of the masses of lightest mesons $M_{\iota,n}(\xi)$ at $\lambda=0.01$  
calculated by semiclassical formulae \eqref{Miot}, \eqref{semicl}:  (a) at $\iota=0$, (b) at $\iota=+$, (c) at $\iota=-$ . \label{fig:Miot}}

\end{figure}

Figure \ref{fig:Miot} displays the evolution of the masses of several lightest meson modes of different parities $\iota=0,\pm$, 
calculated by means of the semiclassical formulas
  \eqref{Miot}, \eqref{semicl}, upon variation of the parameter $\xi$. The parameter $\lambda$ for all 
  mesons in this figure is chosen at the  fixed value $\lambda=0.01$.
  
  As one can see in Figure \ref{fig:M0}, the masses of the mesons with the parity $\iota=0$ monotonically increase
  with decreasing $\xi$. In contrast, the masses of the mesons with positive parity $\iota=+$, 
and fixed $n=1,2,\ldots$, shown in different colours in Figure \ref{fig:M1},  
monotonically decrease with decreasing $\xi$ in the intervals $\xi\in (1/2,\infty)$, 
$\xi\in (1/4,1/2)$,   $\xi\in (1/6,1/4)$,  $\ldots$,  but have discontinuities at the points $\xi=1/2,1/4,1/6,\ldots$.
The masses of the mesons with negative parity $\iota=-$ shown in Figure \ref{fig:M2}  display the similar evolution upon variation of the parameter $\xi$, but with discontinuities at the points $\xi=1,1/3,1/5,\ldots$. 

It is clear, however, that 
the continuity  in $\xi$ of the semiclassical meson mass spectra can be easily restored in both cases $\iota=\pm$ simply by
changing the numbering of the meson states $M_{\iota,n}(\lambda,\xi)$:
\begin{subequations}\label{mesN}
\begin{align}\label{mesNm}
&\tilde{M}_{-,n}(\lambda,\xi)={M}_{-,n-l}(\lambda,\xi),\quad \text{ at } \xi_{2l+1}<\xi<\xi_{2l-1},\\
&\tilde{M}_{+,n}(\lambda,\xi)={M}_{+,n-l}(\lambda,\xi),\quad \text{ at } \xi_{2l+2}<\xi<\xi_{2l},
\end{align}
\end{subequations}
where $\tilde{M}_{\iota,n}(\lambda,\xi)$ denotes the semiclassical meson masses with modified numbering,
$l=0,1,\ldots,n-1$, $\xi_j=1/j$ for natural $j$, and $\xi_0=\xi_{-1}=+\infty$.

\begin{figure}[htb]
\includegraphics[width=\linewidth, angle=00]{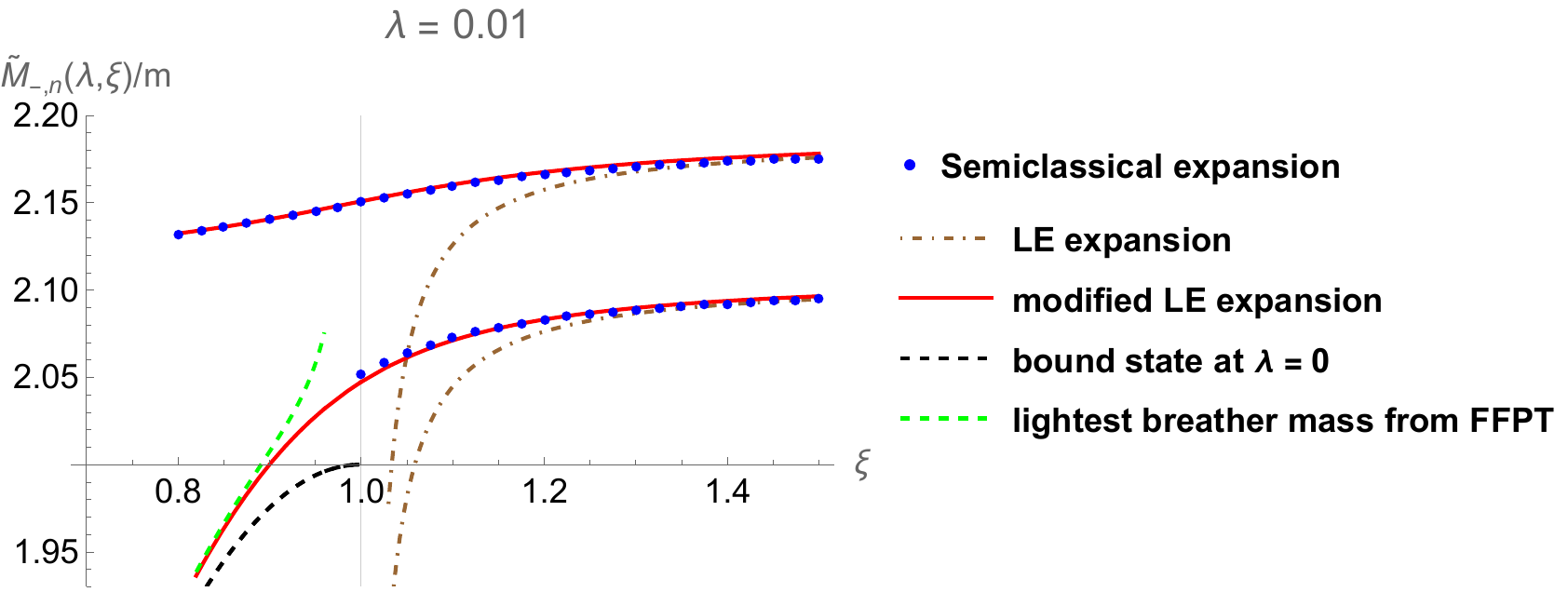}

\caption{Variation with $\xi$ of the masses $\tilde{M}_{-,n}(\lambda,\xi)$ 
 of two lightest meson modes with $n=1,2$ at $\lambda=0.01$ and $\xi$ close to the 
free-fermion point $\xi_1=1$. The $\xi$-dependence of these masses given by the  semiclassical expansion \eqref{Miot}, \eqref{semicl}, \eqref{mesNm}
are shown by blue dots; 
the brown dotted-dashed lines display the variations with $\xi$ of these masses given by the  low-energy expansion \eqref{lE}, \eqref{am}; the solid red lines plot the prediction of the 
modified  low-energy expansion \eqref{Mmin} for $M_{-,n}(0.01,\xi)$ with $n=1,2$. 
The dashed black  line shows the mass of the lightest breather $m_1^{(b)}(\xi)$ at $\lambda=0$, which is   given by \eqref{mbr}. The dashed green line displays the prediction \eqref{br1M}
of the FFPT for the mass of the lightest breather at $\lambda=0.01$.
\label{fig:M-1}} 
\end{figure}

The masses of the meson modes $\tilde{M}_{\iota,n}(\lambda,\xi)$ defined by \eqref{mesN} are continuous (and analytical) functions of the real parameter $\xi$, which monotonically decrease with decreasing $\xi$. However, these functions are defined not for all positive $\xi$:  $\tilde{M}_{-,n}(\lambda,\xi)$ and $\tilde{M}_{+,n}(\lambda,\xi)$  are 
defined in the half-infinite intervals 
  $\xi\in (1/(2n-1),+\infty)$, and  $\xi\in (1/(2n),+\infty)$, respectively. In fact,  the positive 
left edges in these half-infinite intervals indicate the failure in the vicinity of these 
 points of the semi-classical approximation used in the derivation of
 \eqref{semicl}, \eqref{sem2}.  This statement is illustrated in Figure \ref{fig:M-1},  which displays the evolution in 
 $\xi$  close to the point $\xi_1=1$ of two
 lightest meson modes with the negative parity $\iota=-$ at the fixed value of $\lambda=0.01$.  
 The blue dots in this figure plot the $\xi$-dependencies  the masses $\tilde{M}_{-,n}(0.01,\xi)$  with $n=1,2$, 
 determined by the semiclassical formulas \eqref{mesNm}, \eqref{Miot}, \eqref{semicl}. Note, that the explicit form of 
 formula \eqref{mesNm} at  $n=1,2$  reads:
 \begin{align}
&\tilde{M}_{-,1}(\lambda,\xi)={M}_{-,1}(\lambda,\xi), \quad \text{ at } 1<\xi<\infty,\\\label{Sem2a}
&\tilde{M}_{-,2}(\lambda,\xi)=\begin{cases}
{M}_{-,2}(\lambda,\xi), & \text{ at } 1<\xi<\infty,\\
{M}_{-,1}(\lambda,\xi), & \text{ at } 1/3<\xi<1\\
\end{cases}.
\end{align}

 A priori, however, there is no 
 reason to rely on the  semiclassical  asymptotic formulas \eqref{Miot}, \eqref{semicl} in the case of  light mesons with $n=1,2$, 
 since the semiclassical approximation can be well justified only for the highly excited meson states with $n\gg1$. The masses of light mesons should be described, instead, by the low-energy expansions. The brown dotted-dashed curves in Figure  \ref{fig:M-1}
plot at $\xi>1$ the masses $M_{-,n}(0.01,\xi)$ of two lightest mesons (with $n=1,2$) given by the low-energy expansion \eqref{lE}. As one can see from this figure, predictions of the low-energy and semiclassical expansions for these
masses become numerically very close to each other at $\xi\gtrsim1.4$. However, at $\xi$ in the crossover region close to the  point $\xi_1=1$,  the low-energy expansion \eqref{lE} is not applicable, and one should use instead the modified low-energy
expansion  derived in the previous section. The solid red curves in Figure  \ref{fig:M-1} show the predictions of this 
modified low-energy expansion \eqref{Mmin} at $l=1$ for the masses of two lightest mesons ${M}_{-,n}(0.01,\xi)$ with $n=1,2$ at $\xi$ 
in the crossover region close to the point $\xi_1=1$. The   mass  $m_1^{(b)}(\xi)$ of the first breather at $\lambda=0$  
is plotted by the black dashed line in Figure \ref{fig:M-1}. Finally, the mass $m_1^{(b)}(\xi,\lambda)$
of the first breather at $\lambda=0.01$
predicted by the FFTP expansion \eqref{br1M} is shown by the green dashed line. 

Comparison  of the upper blue dotted, and solid red
 curves in Figure \ref{fig:M-1} indicates, that the semiclassical formulas \eqref{Miot}, \eqref{semicl}, \eqref{Sem2a} give  
surprisingly accurate numerical values for the mass   of the second meson mode in the wide interval of the parameter 
$\xi\in (0.8,1.5)$ shown in this figure.
In turn, upon decrease of the parameter $\xi$  below the free-fermion point $\xi_1=1$, the  lower solid red curve approaches 
very close to the dashed green curve, which represents the FFTP prediction  \eqref{br1M} for the 
mass $m_1^{(b)}(\xi,\lambda)$ of the lightest breather. 
This clearly indicates, that the lightest meson having the negative 
parity $\iota=-$ transforms upon decrease of the parameter $\xi$ below the free-fermion point  into the  
\begin{figure}
\centering
\subfloat[ Dashed black  lines display the masses of breathers $m_j^{(b)}(\xi)$ with  $j=2,4,6$ in the unperturbed 
sine-Gordon model. \label{messP}
]
{
\includegraphics[width=.8\linewidth]{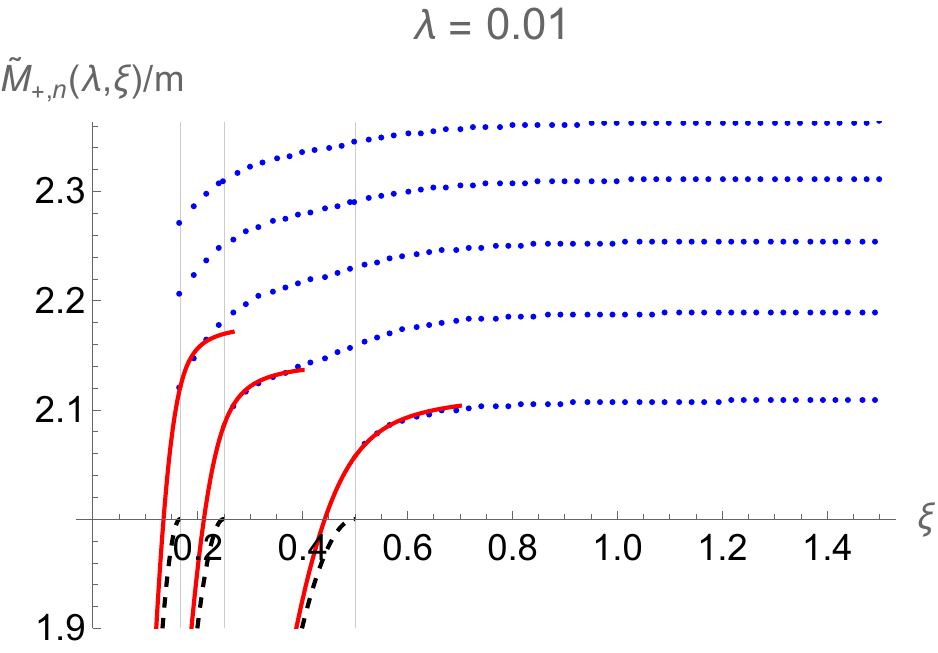}}

\subfloat[ Dashed black  lines display the masses of breathers $m_j^{(b)}(\xi)$ with  $j=1,3,5$ in the unperturbed 
sine-Gordon model.\label{messMa}
]
	{
\includegraphics[width=.8\linewidth]{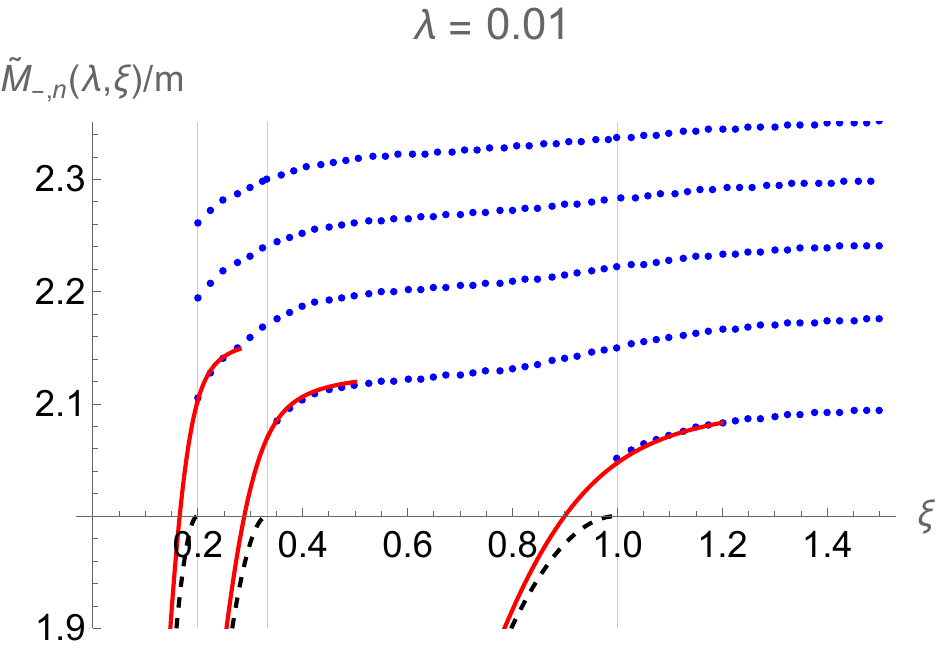}}

\caption{Variation with $\xi$ of the masses of  lightest mesons $\tilde{M}_{\iota,n}(\xi)$ with $n=1,\ldots,5$ at $\lambda=0.01$:  (a) at $\iota=+$, (b) at $\iota=-$.
Blue dots display the semiclassical mass spectra  \eqref{Miot}, \eqref{semicl}. Solid red lines show the  masses of the $n=1$
mesons calculated by means of the modified low-energy expansions:  (a) by formula \eqref{Msp} for the mesons with $\iota=+$; (b)
by formula \eqref{Mmin} for the mesons with $\iota=-$. The masses $m_j^{(b)}(\xi)$ of the  breathers in the unperturbed 
sine-Gordon model, which are shown by black dashed lines, are  given by formula \eqref{mbr}. All masses are normalised by the soliton mass $m$.
\label{fig:Masses3} } 

\end{figure}
breather, deformed  by the perturbation  term $- 2\Lambda \cos(\beta \varphi/2)$ in the interaction potential \eqref{V0}.

It turns out, that all mesons characterized by the parities $\iota=\pm$ display  
qualitatively similar  evolution upon decrease of the parameter $\xi$. Figures \ref{messP}, and \ref{messMa}, 
show the $\xi$-dependencies  of the masses of several meson modes with parities $\iota=+$, and $\iota=-$, respectively.
As in Figures \ref{fig:Miot},  \ref{fig:M-1}, the parameter $\lambda$ is chosen at the  fixed value $\lambda=0.01$. The curves shown 
by the blue dots in Figure \ref{fig:Masses3} display the semiclassical mass spectra determined by equations \eqref{Miot}, 
\eqref{semicl}. The solid red lines show the evolution of the masses $M_{\iota,1}(\lambda,\xi_l+\delta\xi)$ of the lightest mesons, 
which transform into the $l$th breather   in the crossover regions surrounding  the points $\xi_l=1/l$.  The
masses  $M_{\iota,1}(\lambda,\xi_l+\delta\xi)$ in the crossover regions are determined by 
the modified low-energy expansions \eqref{Msp}, and \eqref{Mmin}  for the cases $\iota=+$, and $\iota=-$, respectively. 
The masses of the breathers $m_l^{(b)}(\xi)$ at $\lambda=0$ are shown by the dashed black curves.  
  \section{Conclusions \label{Sec:Conc}}
  In this paper we have analytically studied the particle mass spectrum  in the double  sine-Gordon model
  in the weak confinement regime, that takes place in this model close to the integrable direction, at which the 
  cosine term  with the frequency $\beta /2$ in the scalar self-interaction potential vanishes. This integrability-breaking
  term induces the weak constant attractive force (string tension) between neighbouring solitons leading to their confinement: two solitons bind into  compound particles - the mesons. We obtain three asymptotic expansions in the weak string tension, which describe  the meson masses in different regions of the model parameters. Namely, the semiclassical expansion 
 \eqref{Miot}, \eqref{semicl} describes the masses of higher meson states, while the low-energy expansion \eqref{lE} determines the masses of light mesons at generic values of the coupling constant $\xi$. Finally, at $\xi$ close to  the reflectionless points $\xi_n=1/n$, the masses of light mesons with zero isospin are described by the  modified low-energy expansions 
 \eqref{Mmin} and \eqref{Msp}.
  Analysis of these asymptotic expansions 
  leads us to the unexpected  conclusion, that there is no qualitative difference between the meson and breather excitations
  in the dsG model in the weak confinement regime. Upon increase of  the parameter $\xi$, the $n$th breather does not 
  disappear at the critical point $\xi=1/n$, as it was in the unperturbed sG model, but transforms into  a meson state.

The heuristic perturbative technique, which has been further developed and used in this paper for calculation 
of the meson masses, originates from ideas due to 
McCoy and Wu  \cite{McCoy78}. It is desirable to validate the obtained results
 by reproducing them in the more  systematic  and  powerful approach based on the Bethe-Salpeter equation.
Derivation of the Bethe-Salpeter equation for the dsG field theory would be also crucial for getting 
insight into  the decay mechanism for heavy unstable mesons in this model. 

Of course, it would be interesting to perform the numerical study of the particle mass spectra in the dsG model \eqref{psG0}
by some direct method \cite{TAKACS2006353,Roy2023}, and to  compare the numerical result with our analytical predictions. 
 \section*{Acknowledgments}
I am thankful to Frank G\"ohmann and Sergei Lukyanov for fruitful discussions. This work was supported by Deutsche Forschungsgemeinschaft (DFG) via Grant BO 3401/7-1.
  \appendix
\section{Confinement in the deformed Lieb-Liniger model \label{LibLin}}
The integrable Lieb-Liniger model \cite{Lieb63} describes the system of delta-interacting non-relativistic bosons moving in one-dimension. 
Their state in the $N$-particle sector can be   characterized by the wave function 
$\psi(x_1,x_2,\ldots, x_N)$, 
which remains unchanged under permutations of the spacial coordinates $x_i, x_j$ of any two particles.
In the case of the infinite geometry, the Hamiltonian acting in the $N$-particle sector  reads:
\begin{equation}\label{LL}
H_N=-\sum_{j=1}^N \frac{\partial_{x_j}^2}{2 m}  +\sum_{1\le i<j\le N} c \, \delta(x_i-x_j),
\end{equation} 
where $x_i,x_j\in \mathbb{R}$, and $c$ it the interaction constant.  The free-boson, and free-fermion
cases are realized at $c=0$, and $1/c =+0$ respectively. 

It is well known, that this model admits an alternative equivalent formulation, in which the 
particle coordinates run in the $N$-dimensional region $\Gamma_N$:
\begin{equation}\label{Gam}
\Gamma_N=\{x_1,\ldots,x_N\in \Gamma_N|-\infty<x_1<x_2\ldots<x_N<\infty\}. 
\end{equation}
The wave function $\psi(x_1,x_2,\ldots, x_N)$ must satisfy
the Robin boundary conditions  on $\partial \Gamma_N$: 
\begin{equation}
\lim_{x_{j}\to x_{j+1}}\left[
(\partial_{x_j}-\partial_{x_{j+1}})+c\, m
\right]\psi(x_1,\ldots, x_j,x_{j+1},\ldots,x_N)=0,
\end {equation}
for  $j=1,\ldots,N-1$.
The Hamiltonian acting on such  wave functions   defined in $\Gamma_N$ becomes free:
\begin{equation}\label{HLL}
\widetilde{H}_N=-\sum_{j=1}^N \frac{\partial_{x_j}^2}{2 m}.
\end{equation}
In the two-particle sector, the solution of the Hamiltonian eigenvalue problem 
\begin{equation}
\widetilde{H}_2\,\psi_{p_1,p_2}(x_1,x_1)=\widetilde{E}(p_1,p_2)\, \psi_{p_1,p_2}(x_1,x_1)
\end{equation}
is given by the Bethe wave function
\begin{equation}\label{Beig}
\psi_{p_1,p_2}(x_1,x_1)=\exp[i(p_1 x_1+p_2 x_2)]+\mathcal{S}(p_1-p_2) \exp[i(p_1 x_2+p_2 x_1)],
\end{equation}
where 
\begin{equation}
\mathcal{S}(p_1-p_2)=\frac{p_1-p_2 - i c \, m}{p_1-p_2 + i c \, m}
\end{equation}
is the two-particle scattering amplitude, and 
\begin{equation}\label{pp}
\widetilde{E}(p_1,p_2)=\frac{p_1^2}{2m}+\frac{p_1^2}{2m}
\end{equation}
is the energy of the state \eqref{Beig}.

Besides, a two-particle bound state exists in the attractive regime at $c<0$. Its wave functions reads
\begin{equation}\label{LBound}
\psi_P(x_1,x_2)=\exp\left[\frac{i P(x_1+x_2)}{2}+b\cdot (x_1-x_2)\right],
\end{equation}
where 
\begin{equation}
b= -\frac{c\,m}{2}>0,
\end{equation}
and $P$ is the total momentum of this bound state. 
The energy of the latter equals:
\begin{equation}\label{pE}
\widetilde{E}_1(c,P)=\frac{P^2}{4 m}-\frac{m c^2}{4}.
\end{equation}
\begin{figure}[htb]
\includegraphics[width=\linewidth, angle=00]{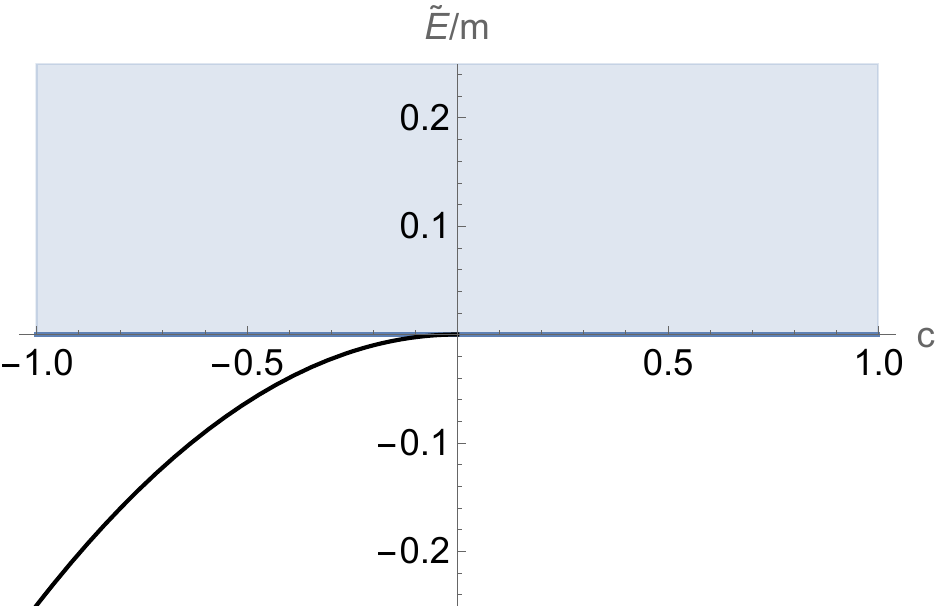}

\caption{The two-particle energy spectra \eqref{pp}, \eqref{pE} of model \eqref{LL} for the states \eqref{Beig}, \eqref{LBound} with zero momentum $P=0.$
\label{fig:Lieb1} } 
\end{figure}
   
Figure \ref{fig:Lieb1} illustrates the variation with the parameter $c$ of the  energy spectra of the Lieb-Liniger model in the two-particle sector for the states with zero total momentum.
The continuous spectrum filling the half-plane $\widetilde{E}>0$ is shown in grey. The parabola $-mc^2/4$ displays the energy \eqref{pE} of the two-particle bound state
 at $P=0$, which exists in  model \eqref{LL}  at $c<0$. 

The model defined by equations \eqref{Gam}-\eqref{HLL} can be viewed as a toy model of a one-dimensional ferromagnet.
Indeed, let us introduce the spin operator $\hat{\sigma}(x)$, which acts  on the wave-function 
$\psi(x_1,x_2,\ldots, x_N)$
defined in $\Gamma_N$  as follows:
\begin{equation}
\hat{\sigma}(x)\, \psi(x_1,x_2,\ldots, x_N) =\sigma(x|x_1,x_2\ldots,x_N)\, \psi(x_1,x_2,\ldots, x_N).
\end{equation}
The spin variable $\sigma(x|x_1,x_2\ldots,x_N)$ takes values $\pm \bar{\sigma}$ for $x$ in the interval  $(x_j,x_{j+1})$ bounded by the 
particle coordinates:
\begin{equation}
\sigma(x|x_1,x_2\ldots,x_N)=(-1)^{j}\bar{\sigma}, \text{ for } x\in (x_j,x_{j+1}),
\end{equation}
with $j=0,1,2,\ldots,N$, and $x_0=-\infty$, $x_{N+1}=+\infty$. In this interpretation, the particle coordinates $x_j$ 
are treated as
locations of  kinks separating the oppositely magnetized ferromagnetic domains. 

Let us now deform the Lieb-Liniger model \eqref{Gam}-\eqref{HLL} by application of the uniform magnetic field $h$ coupled to the
spin operator $\hat{\sigma}$:
\begin{equation}\label{HNh}
\widetilde{H}_N(h)=-\sum_{j=1}^N \frac{\partial_{x_j}^2}{2 m}-h \int_{-\infty}^\infty dx\, [\hat{\sigma}(x)-\bar{\sigma}].
\end{equation}
The deformation term breaks integrability of the model and leads to the confinement of kinks. However, in contrast to
other QFTs (Ising field theory, Potts field theory, sine-Gordon model, etc.), the deformation term in the Hamiltonian \eqref{HNh}
commutes with the operator of number of particles. Furthermore, the model \eqref{HNh}
remains integrable in the two-particle sector $N=2$. The corresponding stationary Schr\"odinger equation reads:
\begin{equation}\label{eqAi}
-\frac{\partial_{x_1}^2 \psi(x_1,x_2)}{2m}-\frac{\partial_{x_2}^2 \psi(x_1,x_2)}{2m}+ f \,(x_2-x_1)\, \psi(x_1,x_2)=\widetilde{E}\, \psi(x_1,x_2),
\end{equation}
where $f=2 h \bar{\sigma}$ is the string tension. The two-particle wave function $\psi(x_1,x_2)$ is defined in the half-plane $-\infty<x_1<x_2<\infty$,
and satisfies in the line $x_1=x_2$ the  boundary condition:
\begin{equation}
\lim_{x_1\to x_2}(\partial_{x_1}-\partial_{x_2}+ c \,m ) \,\psi(x_1,x_2)=0.
\end{equation}
For the translation invariant eigenfunction $\psi(x_1,x_2)=\psi(x_1+X, x_2+X)$,  with an arbitrary real $X$,
one  obtains from \eqref{eqAi} the ordinary differential equation  of the Airy type:
\begin{equation}\label{AirA}
- \frac{\phi''(x)}{m}- f \,x\, \phi(x)=\widetilde{E}\, \phi(x),
\end{equation}
where $x=x_1-x_2<0$, and $\phi(x)=\psi(x,0)$.

The solution  of this equation vanishing  at $x\to-\infty$ reads: 
\begin{equation}
\phi(x)=\mathrm{Ai}\left(-\lambda^{1/3} m\, x- \frac{\widetilde{E}}{\lambda^{2/3}m}\right),
\end{equation}
where $\lambda=f/m^2$.

The boundary condition 
\begin{equation}\label{BC2}
\lim_{x\to 0-} [2\phi'(x)+ c m \,\phi(x)]=0,
\end{equation}
leads to the transcendent secular equation
\begin{equation}\label{Yro}
Y_\rho(t_n)=0,
\end{equation}
where 
\begin{equation}
Y_\rho(t)=\mathrm{Ai}'(-t)-\rho \,\mathrm{Ai}(-t),
\end{equation}
and
\begin{equation}\label{rh}
\rho=\frac{c}{2 \lambda^{1/3}}.
\end{equation}
Solutions  $\{t_n\}_{n=1}^\infty$ of  equation \eqref{Yro} determine the energy spectrum $\{\widetilde{E}_n\}_{n=1}^\infty$  of the boundary problem \eqref{AirA},
\eqref{BC2}:
\begin{equation}\label{Ent}
\widetilde{E}_n=m \lambda^{2/3} t_n.
\end{equation}

\begin{figure}
\centering
\subfloat[ 
]
{\label{et1}
\includegraphics[width=.5\linewidth]{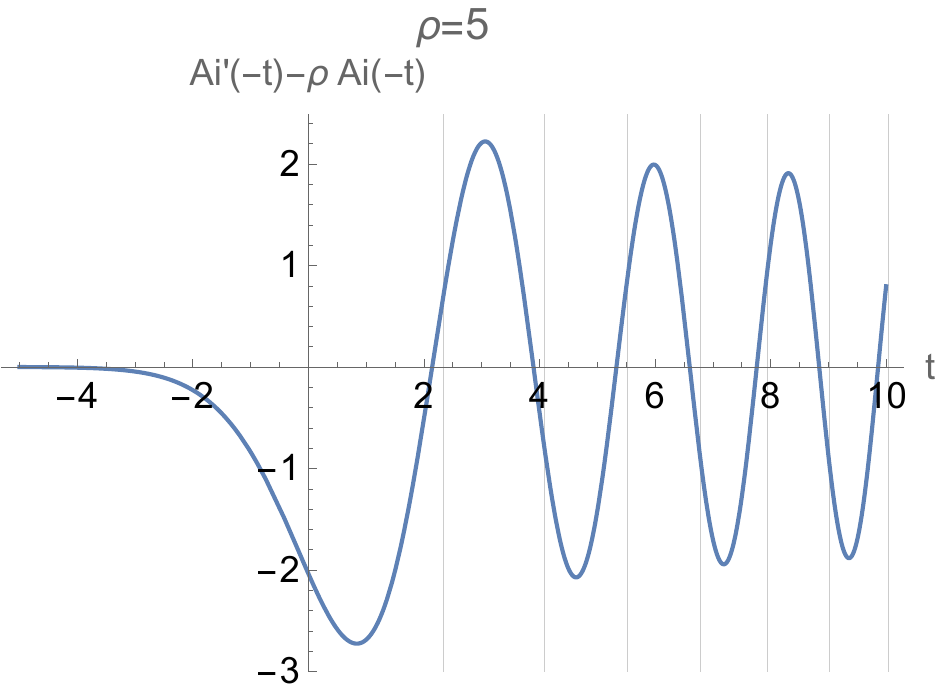}}

\subfloat[
]
	{\label{et2}
\includegraphics[width=.5\linewidth]{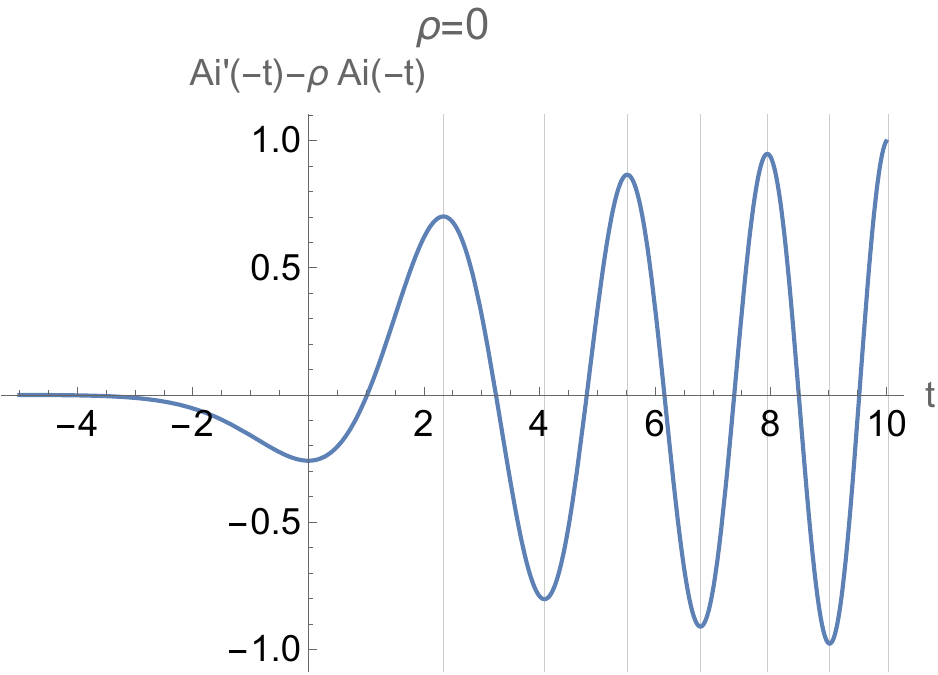}}

\subfloat[ 
]
{\label{et1a}
\includegraphics[width=.5\linewidth]{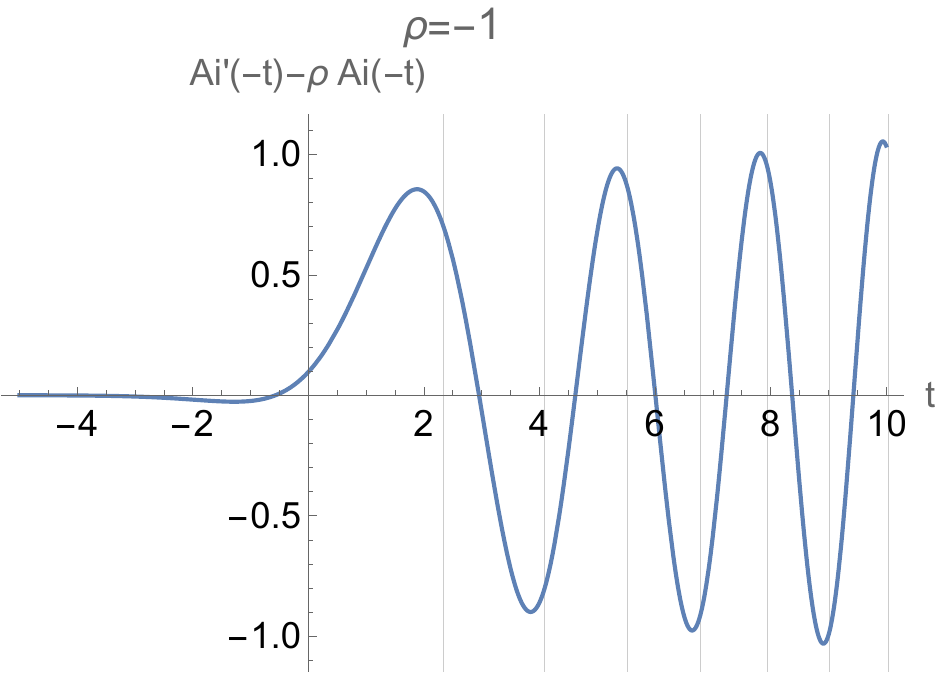}}

\caption{Plots of the function $Y_\rho(t)=\mathrm{Ai}'(-t)-\rho \,\mathrm{Ai}(-t)$ at three different values of the 
parameter $\rho$: (a) $\rho=5$; (b) $\rho=0$;  (c) $\rho=-1$. Vertical lines indicate the zeroes $\{z_n\}_{n=1}^6$ of the 
function $\mathrm{Ai}(-z)$.\label{fig:Ai}}

\end{figure}

The limit $\rho\to+\infty$ corresponds to the free-fermionic case. In this limit, the numbers $ -t_n$ approach to the zeroes of the Airy function:
\[
\lim_{\rho\to +\infty} t_n=z_n, \quad \mathrm{Ai}(-z_n)=0.
\]

At $\rho=0$ the free-bosonic case is realized, which is illustrated in Figure~\ref{fig:Ai}b. In this case the 
numbers $ -t_n$ coincide with the zeroes of the derivative of the Airy function: 
\[
\lim_{\rho\to 0} t_n=z_n', \quad \mathrm{Ai}'(-z_n')=0.
\]

The plot of the function $Y_\rho(t)$ at $\rho=-1$ is shown in Figure~\ref{fig:Ai}c. The first zero 
of this function is negative
 $t_1(-1)=-0.566891$, while all other zeroes   remain positive.
 
 \begin{figure}[htb]
\includegraphics[width=\linewidth, angle=00]{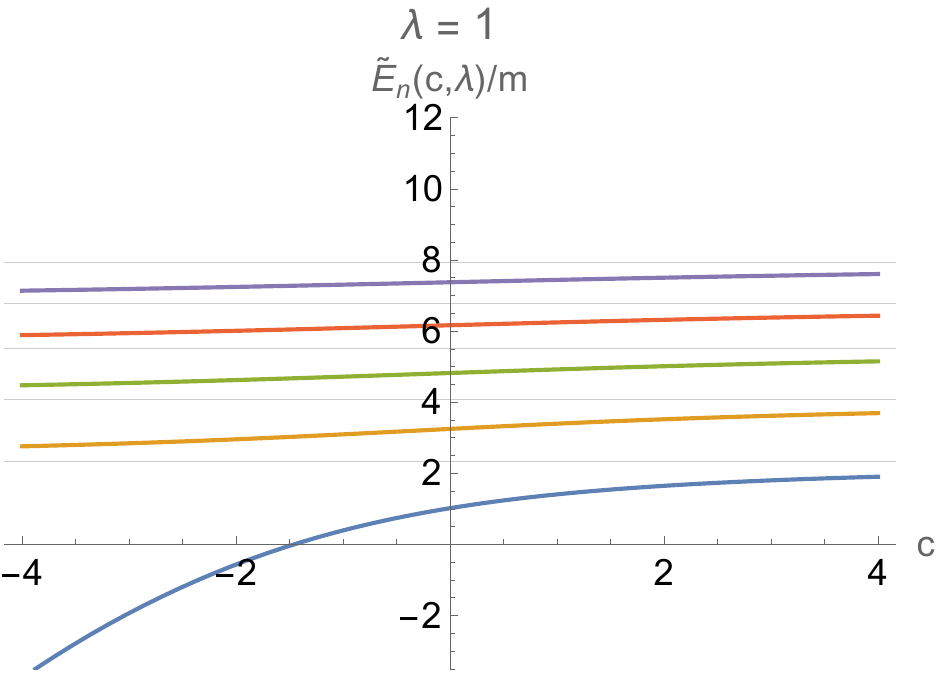}

\caption{Solid lines display variation with the parameter $c$ of the energies \eqref{Ent} of the mesons with zero momentum $P=0$
in model \eqref{HNh} at $\lambda=1$. The dashed black parabolic line shows the energy of the two-particle bound  state at $\lambda=0$.
Horizontal lines are located at $\lambda^{2/3}z_n$, with $n=1\ldots,5$.
\label{fig:Lieb2} } 
\end{figure}

Finally,  at $\rho\to-\infty$, one gets from equations \eqref{rh}, \eqref{Ent}:
\begin{equation}
\widetilde{E}_1=-\frac{m c^2}{4}, \quad \widetilde{E}_2=m \lambda^{2/3} z_1, \quad \widetilde{E}_3=m \lambda^{2/3} z_2, \quad \ldots.
\end{equation}

Figure \ref{fig:Lieb2} shows the plots of the meson energies $\widetilde{E}_n(c,\lambda)$ given by equation \eqref{Ent} versus the parameter $c$ at $\lambda=1$.

The energy spectrum \eqref{Ent} has the following properties.
\begin{enumerate}
\item At any fixed $c\in \mathbb{R}$ and $\lambda>0$, the energy spectrum is discrete. 
\item At $\lambda>0$, all meson energies $\widetilde{E}_n(c,\lambda)$, with $n=1,2,\ldots$, analytically depend on $c$ at $-\infty<c<\infty$.
\item
At $\widetilde{E}>0$, the spectrum becomes more and more dense with decreasing $\lambda$,
and transforms into the continuous spectrum at $\lambda=0$.
\item 
The energy of the lightest meson $\widetilde{E}_1(c,\lambda)$ becomes negative at 
$c<c_0(\lambda)$, where $c_0(\lambda)=2  \lambda^{1/3}\, \mathrm{Ai}'(0)/\mathrm{Ai}(0)\approx-1.4580\, \lambda^{1/3} $.
\item 
At $c<0$, the lightest meson  transforms in the limit $\lambda\to+0$ into the bound state
of the unperturbed Lieb-Liniger model \eqref{LL}, and 
\[
\lim_{\lambda\to+0}\widetilde{E}_1(c,\lambda)=-\frac{m\,c^2}{4}, \text{ at }c<0.
\]
\end{enumerate}

\end{document}